\documentclass{myelsart}
\usepackage{ifpdf}
\usepackage{csquotes}
\usepackage{caption}
\usepackage{graphicx}
\usepackage{amsmath}
\usepackage{epsf}
\usepackage{graphpap}
\usepackage{placeins}
\usepackage{xcolor}
\usepackage{subfig}
\usepackage{relsize}
\usepackage{caption}
\usepackage{array} 
\usepackage{multirow} 
\ifpdf
\usepackage[%
  pdftitle={Instructions for use of the document class
    elsart},%
  pdfauthor={Simon Pepping},%
  pdfsubject={The preprint document class elsart},%
  pdfkeywords={instructions for use, elsart, document class},%
  pdfstartview=FitH,%
  bookmarks=true,%
  bookmarksopen=true,%
  breaklinks=true,%
  colorlinks=true,%
  linkcolor=blue,anchorcolor=blue,%
  citecolor=blue,filecolor=blue,%
  menucolor=blue,pagecolor=blue,%
  urlcolor=blue]{hyperref}
\else
\usepackage[%
  breaklinks=true,%
  colorlinks=true,%
  linkcolor=blue,anchorcolor=blue,%
  citecolor=blue,filecolor=blue,%
  menucolor=blue,pagecolor=blue,%
  urlcolor=blue]{hyperref}
\fi

\makeatletter
\def\elsartstyle{%
    \def\normalsize{\@setfontsize\normalsize\@xiipt{14.5}}
    \def\small{\@setfontsize\small\@xipt{13.6}}
    \let\footnotesize=\small
    \def\large{\@setfontsize\large\@xivpt{18}}
    \def\Large{\@setfontsize\Large\@xviipt{22}}
    \skip\@mpfootins = 18\p@ \@plus 2\p@
    \normalsize
} \@ifundefined{square}{}{\let\Box\square} \makeatother

\usepackage{amstext}
\usepackage{amssymb}
\usepackage{graphicx}
\usepackage{amsmath}
\usepackage{epsf}
\usepackage{graphpap}
\usepackage{xcolor}
\newcommand{\be}{\begin{equation}}
    \newcommand{\ee}{\end{equation}}
\newcommand{\ba}{\begin{array}{c}}
    \newcommand{\ea}{\end{array}}
\def\eqref#1{(\ref{#1})}

\usepackage{xcolor}
\usepackage{mathtools}
\usepackage{tensor}
\definecolor{Red}{rgb}{0.9,0.2,0.1}
\definecolor{dkRed}{rgb}{0.5,0.2,0.4}
\definecolor{Green}{rgb}{0.2,0.9,0.2}
\definecolor{dkGreen}{rgb}{0.1,0.8,0.1}
\definecolor{Yellow}{rgb}{1,1,0}
\definecolor{Navy}{rgb}{0.1,0.1,0.4}
\definecolor{Navy2}{rgb}{0.1,0.1,0.5}
\definecolor{Black}{rgb}{0,0,0}
\definecolor{Orange}{rgb}{1,0.55,0.02}
\definecolor{Pink}{rgb}{0.88,0.09,0.77}
\definecolor{Grey}{rgb}{0.7,0.7,0.7}
\definecolor{Math}{rgb}{0.07,0.63,0.08}
\definecolor{Violet}{rgb}{0.4,0,0.7}

\begin{document}
\begin{frontmatter}

\title{Lie symmetries, closed-form solutions, and conservation laws of a constitutive equation modeling stress in elastic materials}
\author{Rehana Naz$^{1*}$, Willy Hereman$^{2}$}

\address{$^{1}$Department of Mathematics and Statistical Sciences, Lahore School of Economics, Lahore 53200, Pakistan
\\
Department of Applied Mathematics and Statistics,
Colorado School of Mines,
Golden CO 80401-1887, USA
\\
$^{*}$Corresponding author: drrehana@lahoreschool.edu.pk}

\begin{abstract}
The Lie-point symmetry method is used to find some closed-form
solutions for a constitutive equation modeling stress in elastic
materials. The partial differential equation (PDE), which involves a
power law with arbitrary exponent $n$, was investigated by Mason and
his collaborators (Magan {\em et al.}, Wave Motion, 77, 156-185,
2018). The Lie algebra for the model is five-dimensional for the
shearing exponent $n>0$, and it includes translations in time,
space, and displacement, as well as time-dependent changes in
displacement and a scaling symmetry. Applying Lie's symmetry method,
we compute the optimal system of one-dimensional subalgebras. Using
the subalgebras, several reductions and closed-form solutions for
the model are obtained both for arbitrary exponent $n$ and special
case $n = 1$. Furthermore, it is shown that for arbitrary $n>0$ the
model has interesting conservation laws which are computed with
symbolic software using the scaling symmetry of the given PDE.
\end{abstract}
\end{frontmatter}
\vspace*{-2mm}
\section{Introduction}
\label{intro}
\vspace*{-2mm}
In this paper we perform a Lie symmetry analysis, compute
closed-form solutions, and conservation laws of a constitutive equation investigated by
Kannan {\em et al.} \cite{kannan-rajagopal-saccomandi-wave-2014}
and Mason {\em et al.} \cite{magan-etal-wm-2018}.
In non-dimensional form the governing partial differential
equations (PDEs), which model stress and displacement in
elastic materials, read
\begin{eqnarray}
\label{magan-(3.7)-(3.8)}
\sigma_y = \delta u_{tt}, \quad u_y
= \tfrac{1}{\delta} \sigma (\beta + \sigma^2)^n,
\end{eqnarray}
where $\sigma(y,t)$ is the shear stress and $u(y,t)$ is
displacement. Furthermore, $y$ and $t$ are a spatial variable and
time, respectively, and subscripts denote partial derivatives, e.g.,
$u_{tt} = \tfrac{\partial^2 u}{\partial t^2}$. Parameter $\delta$ is
a real constant and $n \ge 0$ is a shearing exponent which can be an
integer or rational number. An auxiliary constant parameter $\beta$
has been introduced\footnote{One can set $\beta = 1$ in the results
of the computations.} to make the system scaling homogeneous as
explained in Section~\ref{cons-laws}.
The reciprocal of $\delta$, i.e., $\tfrac{1}{\delta} =
\frac{\alpha}{\sqrt{\gamma}}$, is the displacement gradient which
involves two material parameters $\alpha$ and $\gamma$.
We do not assume that the displacement gradient is small which would allow one to find solutions with a perturbation method as was done in \cite{magan-etal-wm-2018}.
We also exclude the case $n=0$ because we focus on {\em nonlinear}
models of type (\ref{magan-(3.7)-(3.8)}).

The dependent variable $u$ (and simultaneously parameter $\delta$)
can be eliminated by replacing (\ref{magan-(3.7)-(3.8)}) with
\begin{equation}
\label{magan-(3.9)}
\sigma_{yy} = \left( \sigma (\beta + \sigma^2)^n \right)_{tt},
\end{equation}
which is a single hyperbolic PDE describing shear stress waves.
Due to the presence of an arbitrary exponent $n$,
Magan \cite {magan-thesis-2017} called
(\ref{magan-(3.7)-(3.8)}) a power-law constitutive
equation due to its analogy with constitutive equations in fluid dynamics (see, e.g.,
\cite{magan-etal-ijnm-2016,magan-etal-ijnm-2017}).
Therefore, the methodology used in our paper also applies to fluid dynamics as well as to {\em nonlinear} wave equations of type (\ref{magan-(3.9)}) wherever they arise.

We use Lie group methods to establish closed-form solutions and some
conservation laws for the coupled non-linear PDEs
(\ref{magan-(3.7)-(3.8)}). More precisely, we compute the optimal
system of one-dimensional subalgebras. Using the subalgebras,
several reductions and closed-form solutions for the model are
obtained both for arbitrary $n$ and special case $n = 1$. In
contrast to asymptotic solutions of (\ref{magan-(3.9)}) and
approximate standing and traveling wave solutions of
(\ref{magan-(3.7)-(3.8)}) computed in \cite{magan-etal-wm-2018} with
standard perturbation methods, the Lie symmetry method leads to
exact solutions of (\ref{magan-(3.7)-(3.8)}) for which the physical
relevance has not been investigated. The derivation of closed-form
solutions through Lie group methods is vast and well-established in
the academic literature. Seminal books on the subject include
\cite{bluman-kumei-springer-1989,bluman-etal-book-2010,
ibragimov-crc-1994,ovsiannikov-academic-1982,olver-springer-1993}.
Several symbolic packages are developed to derive the Lie symmetries
and handle various tasks related to Lie group methods
\cite{butcher-etal-cpc-2003,carminati-vu-jsc-2000,champagne-etal-1991,
chebterrab-von-bulow-cpc-1995,cheviakov-2007,
goktas-hereman-code-invarsym-1997,hereman-mcm-1997,
hereman-goktas-mca-2024,rocha-figueiredo-cpc-2011,
vu-butcher-carminati-cpc-2007}.

To compute conservation laws, we use a direct approach based on
the scaling symmetry of the original PDE system.
As far we know, the conversation laws we have found are new and the most complicated ones might be hard to compute with the multiplier method
\cite{cheviakov-zhao-book-2024,mubai-mason-symmetry-2022,
naz-mahomed-mason-amc-2008}
or partial Lagrangian technique
\cite{naz-mahomed-mason-narwa-10(6)-2009}.
Regardless of the method used, computing conservation laws is a
non-trivial matter, in particular, for systems involving an
arbitrary exponent $(n)$. The computations could likely not have
been done without the use of specialized symbolic software packages
such as \verb|ConservationLawsMD.m|
\cite{poole-hereman-code-conslawsmd-2009} developed by Poole and
Hereman \cite{poole-hereman-jsc-2011}, Cheviakov's {\em Maple} code
\verb|GeM|
\cite{bluman-etal-book-2010,cheviakov-2007,cheviakov-2014}, and the
{\em Maple} based package \verb|SADE|
\cite{rocha-figueiredo-cpc-2011}.

To date, even the most sophisticated codes only work for systems where all variables have fixed exponents.
Therefore, some interactive work, insight, and ingenuity are required to find conservation laws for parameter-dependent systems,
in particular, those with arbitrary exponents.
Once the general forms of the densities and fluxes
are established, testing that they satisfy the conservation law
is straightforward but can be cumbersome and is prone to errors if done by hand.
All conservation laws presented in this paper were verified independently with \verb|ConservationLawsMD.m| \cite{poole-hereman-code-conslawsmd-2009}
and the \verb|ConservedCurrentTest| option of the package \verb|PDETools|
developed by Cheb-Terrab and von B\"ulow
\cite{chebterrab-von-bulow-cpc-1995}, now built into {\em Maple}.

The research presented in this paper is very much in the spirit of some of the work that Mason has done throughout his illustrious career.
Using the Lie symmetry method, Mason and his collaborators have derived closed-form solutions and conservation laws of numerous
differential equations (some involving power laws) arising in mechanics and fluid mechanics
(see, e.g.,
\cite{magan-thesis-2017, magan-etal-ijnm-2016,fareo-mason-cnsns-2013,fareo-mason-ijmpb-2016,
hutchinson-mason-mahomed-cnsns-2015}.
However, the method has also been successfully applied to
mathematical models in, e.g.,
economics, epidemiology, and other areas of applied mathematics  \cite{chaudhry-naz-dcdss-2024,
freire-torrisi-cnsns-2014, irum-naeem-mmas-2024}.

The paper is organized as follows. The Lie-point symmetry generators
for (\ref{magan-(3.7)-(3.8)}) are computed in
Section~\ref{Lie-point-symm}. In Section~\ref{optical-system}, the
optimal system of one-dimensional subalgebras is derived. Using
these subalgebras, in Sections~\ref{closed-form-sols-Y2} and
\ref{traveling-waves} the PDEs are reduced to ODEs for which
closed-form solutions are computed. Section~\ref{cons-laws} covers
the computation of conservation laws using the scaling homogeneity
approach. The results are briefly discussed in
Section~\ref{conclusions-future-work} where also a few topics for
future work are mentioned. Finally, in Section~\ref{dedication} the
authors express their gratitude to Prof.\ David Mason.
\vspace*{-2mm}
\section{Lie-point symmetries for system (\ref{magan-(3.7)-(3.8)})} \label{Lie-point-symm}
\vspace*{-2mm}
In this section we compute the Lie-point symmetries of
(\ref{magan-(3.7)-(3.8)}). To perform a Lie symmetry
analysis \cite{bluman-kumei-springer-1989,bluman-etal-book-2010,
ibragimov-crc-1994,ovsiannikov-academic-1982,olver-springer-1993}
we write the system as
\begin{equation}
\label{lie-magan-(3.7)-(3.8)}
E^1(y,t,\sigma,u,\sigma_{y},u_{tt}) =
0, \;\;
E^2(y,t,\sigma,u,u_{y}) = 0,
\end{equation}
where
\begin{equation}
\label{alt-lie-magan-(3.7)-(3.8)}
E^1 = \sigma_y -\delta u_{tt}, \;\;
E^2 = u_y - \tfrac{1}{\delta} \sigma (1 + \sigma^2)^n,
\end{equation}
after setting $\beta=1$.

A symmetry infinitesimal generator for
(\ref{magan-(3.7)-(3.8)}) is of the form
  \begin{equation}
  \label{gen-symm}
    X =
    \xi^1(y,t,\sigma, u)\frac{\partial }{\partial y}
    + \xi^2(y,t,\sigma, u)\frac{\partial }{\partial t}
    + \eta^1(y,t,\sigma, u)\frac{\partial }{\partial \sigma}
    + \eta^2(y,t,\sigma, u)\frac{\partial }{\partial u},
  \end{equation}
  and is derived from the following equations
  (also known as the invariance conditions \cite{bluman-kumei-springer-1989,bluman-etal-book-2010,
  ibragimov-crc-1994,ovsiannikov-academic-1982,olver-springer-1993}):
  \begin{eqnarray}
  \label{inv-symm}
    \left\{
    \begin{array}{l}
        X^{[2]}E^1\mid_{(E^1=0,\;E^2=0)}\, = \, 0,
        \\
       X^{[2]}E^2\mid_{(E^1=0,\;E^2=0)}\, = \,0,
    \end{array}
    \right.
  \end{eqnarray}
where $X^{[2]}$ is the second prolongation of generator $X$, given
by
\begin{eqnarray}
\label{prolong-X} X^{[2]} &=& X + \zeta^1_y\frac{\partial }{\partial
\sigma_y} + \zeta^2_y\frac{\partial }{\partial u_y}+
\zeta^1_t\frac{\partial }{\partial \sigma_t} +
\zeta^2_t\frac{\partial }{\partial u_t}
\nonumber \\
&&+ \zeta^1_{yy}\frac{\partial }{\partial \sigma_{yy}}  +
\zeta^2_{yy}\frac{\partial }{\partial u_{yy}}+
\zeta^1_{yt}\frac{\partial }{\partial \sigma_{yt}} +
\zeta^2_{yt}\frac{\partial }{\partial u_{yt}}+
\zeta^1_{tt}\frac{\partial }{\partial \sigma_{tt}} +
\zeta^2_{tt}\frac{\partial }{\partial u_{tt}}  .
\end{eqnarray}
As usual, the expressions of the coordinates
$\zeta^1_{y}$, $\zeta^1_{t}$, $\zeta^1_{yy}$,$ \cdots$, $\zeta^2_y$,
etc., are written as
(see, e.g., \cite{bluman-kumei-springer-1989,bluman-etal-book-2010,
ibragimov-crc-1994,ovsiannikov-academic-1982,olver-springer-1993})
 \begin{eqnarray}
\!\!\!\!\!\!\!\!\!\!
 \zeta^1_y
 &=&
 D_y(\eta^1 )-\sigma_yD_y(\xi^1)-\sigma_tD_y(\xi^2),
 \;\;
 \zeta^1_t =
 D_t(\eta^1 )-\sigma_yD_t(\xi^1)-\sigma_tD_t(\xi^2),
 \nonumber\\
 \!\!\!\!\!\!\!\!\!\!
 \zeta^2_y
 &=&
 D_y(\eta^2 )-u_yD_y(\xi^1)-u_tD_y(\xi^2),
 \;\;
 \zeta^2_t =
 D_t(\eta^2 )-u_yD_t(\xi^1)-u_tD_t(\xi^2),
 \nonumber \\
 \!\!\!\!\!\!\!\!\!\!
\zeta^1_{yy}
 &=&
 D_y(\zeta^1_y)-\!\sigma_{yy}D_y(\xi^1)-\!\sigma_{yt}D_y(\xi^2),
 \;\;
 \zeta^2_{yy} =
 D_y(\zeta^2_y)-\!u_{yy}D_y(\xi^1)-\!u_{yt}D_y(\xi^2),
 \nonumber \\
 \!\!\!\!\!\!\!\!\!\!
 \zeta^1_{ty}
 &=&
 D_y(\zeta^1_t)-\!\sigma_{yt}D_y(\xi^1)-\!\sigma_{tt}D_y(\xi^2),
 \;\;
 \zeta^2_{ty} =
 D_y(\zeta^2_t)-\!u_{yt}D_y(\xi^1)-\!u_{tt}D_y(\xi^2),
\nonumber\\
\!\!\!\!\!\!\!\!\!\!
 \zeta^1_{tt}
 &=&
 D_t(\zeta^1_t)-\!\sigma_{yt}D_t(\xi^1)-\!\sigma_{tt}D_t(\xi^2),
 \;\;
 \zeta^2_{tt} =
 D_t(\zeta^2_t )-\!u_{yt}D_t(\xi^1)-\!u_{tt}D_t(\xi^2).
 \label{formulas-zeta}
 \end{eqnarray}
The total derivative operators $D_y$ and $D_t$ are defined by
\begin{eqnarray}
\label{Dy-operator} \!\!\ \!\! \!\!\ \!\! \!\!\ \!\! D_y =
 \frac{\partial}{ \partial y}
 + \sigma_y \frac{\partial} {\partial \sigma}
 + u_y \frac{\partial} {\partial u}
 + \sigma_{yy} \frac{\partial} {\partial \sigma_y}
 + u_{yy} \frac{\partial} {\partial u_y}
 + \sigma_{yt} \frac{\partial}{\partial \sigma_t}
 + u_{yt} \frac{\partial}{\partial u_t} + \ldots.
 \end{eqnarray}
 and
\begin{eqnarray}
\label{Dt-operator}
 \!\!\ \!\! \!\!\ \!\! \!\!\ \!\!
 D_t =
 \frac{\partial}{ \partial t}
 + \sigma_t \frac{\partial} {\partial \sigma}
 + u_t \frac{\partial} {\partial u}+\sigma_{tt} \frac{\partial} {\partial \sigma_t}
 + u_{tt} \frac{\partial} {\partial u_t}
 + \sigma_{yt} \frac{\partial}{\partial \sigma_y}
 + u_{yt} \frac{\partial}{\partial u_y}
 + \ldots,
 \end{eqnarray}
The system {(\ref{inv-symm})} is separated according to the
derivatives of $\sigma$ and $u$
to yield an overdetermined system of
 {\em linear} PDEs for the unknown coefficients
 $\xi^1$, $\xi^2$, $\eta^1$ and $\eta^2$.
 More efficiently than doing it by hand, the determining equations for the Lie-point symmetries can be computed with
 symbolic software packages \cite{butcher-etal-cpc-2003,carminati-vu-jsc-2000,hereman-mcm-1997}
 such as
 \cite{champagne-etal-1991,cheviakov-2007,rocha-figueiredo-cpc-2011}.
We used {\em Maple}-based package \verb|DESOLV II|
\cite{vu-butcher-carminati-cpc-2007},
which works for arbitrary $n$, to get
\begin{eqnarray}
  \label{Liesym-det-eq}
\!\!\!\!\!\!\!\!
&& \xi^1_t=0, \;\;\; \xi^1_{\sigma}=0,
\;\;\; \xi^1_u = 0, \;\;\;
\label{Liesym-det-eq1}
\xi^2_y = 0, \;\;\; \xi^2_{\sigma} = 0, \;\;\; \xi^2_u = 0,
\\
\!\!\!\!\!\!\!\!
&& \xi^2_{tt}-2 \eta^2_{ut}=0, \;\;\;
\eta^2_{\sigma} = 0, \;\; \eta^2_{uu} = 0, \;\;\;
    2 \xi^2_t-\xi^1_y+\eta^1_{\sigma}-\eta^2_{u} = 0,
    \label{Liesym-det-eq4}
\\
\!\!\!\!\!\!\!\!
&& \sigma (1 + \sigma^2)^n\eta^1_u
    + \delta \eta^1_y-\delta^2 \eta^2_{tt} = 0,
    \label{Liesym-det-eq5}
    \\
\!\!\!\!\!\!\!\!
  &&\sigma(1 + \sigma^2)^{n+1}(\xi^1_y-\eta^2_u)
  +\! (1 + (2n+1)\sigma^2) (1+\sigma^2)^n  \eta^1
  -\!\delta (1 + \sigma^2)\eta^2_y = 0.
  \label{Liesym-det-eq6}
  \end{eqnarray}
System \eqref{Liesym-det-eq1}-\eqref{Liesym-det-eq4} can be solved
straightforwardly (by hand or with symbolic software) yielding the
following general solution:
\begin{eqnarray}
\label{solve-det-eqn-general}
  \xi^1 &=& G_1(y), \;\;
  \xi^2 = G_2(t),
  \nonumber \\
  \eta^1 &=&
  \left( G_1'(y)-\tfrac{3}{2}G_2'(t)
  + G_3(y) \right)\sigma +G_5(y,t,u),
  \\
  \eta^2 &=& \left( \tfrac{1}{2}G_2'(t) + G_3(y) \right) u
  + G_4(y,t),
  \nonumber
  \end{eqnarray}
where $G_1(y),\;G_2(t),\;G_3(y),\;G_4(y,t)$ and  $G_5(y,t,u)$ are
arbitrary functions.
The remaining equations, i.e., \eqref{Liesym-det-eq5} and
\eqref{Liesym-det-eq6}, then take the following form
 \begin{eqnarray}
 \label{Liesym-det-eq5a}
  && G_{5u}(y,t, u)  (1 + \sigma^2)^n \sigma
   + \delta \big( G_1''(y) + G_3'(y) \big) \sigma
   -  \tfrac{1}{2} \delta^2 G_2'''(t)  u
 \nonumber \\
 &&\; + \delta G_{5y}(y,t, u)  - \delta^2 G_{4tt}(y,t) =0,
 \end{eqnarray}
 and
  \begin{eqnarray}
 \label{Liesym-det-eq6a}
  && \big(2n G_3(y)
 + 2(n+1)G_1'(y)-(3n+2)G_2'(t)\big) \sigma^3(1 + \sigma^2)^n
 \nonumber\\
 &&\; + \big((2n+1)G_5(y,t,u)\sigma^2 + 2(G_1'(y)-G_2'(t))\sigma
 + G_5(y,t,u)\big) ( 1 + \sigma^2)^n
 \nonumber\\
 && \; -\delta \big(
  G_{4y}(y,t)\sigma^2 + G_3'(y)\sigma^2 u
  + G_3'(y)u + G_{4y}(y,t) \big) = 0.
\end{eqnarray}
When \eqref{Liesym-det-eq6a} is separated with respect to
$\sigma^2 (1 + \sigma^2)^n$, where $n$ is a natural number, it yields
$G_5=0$.
After substituting $G_5=0$ into equations \eqref{Liesym-det-eq5a}
and \eqref{Liesym-det-eq6a}, it becomes possible to separate those
with respect to different combinations of $u$ and $\sigma$, allowing
one to find
$G_1(y)$, $G_2(t)$, $G_3(y)$, and $G_4(y,t)$. Indeed,
\eqref{Liesym-det-eq5a} and \eqref{Liesym-det-eq6a} lead to
 \begin{eqnarray}
 && G_1''(y) + G_3'(y) = 0,
 \;\;
 G_2'''(t) = 0,
 \;\;
 G_{4tt}(y,t) = 0,
 \label{Liesym-det-eq-simplified1}
 \\
&& n \left(G_3(y) + G_1'(y)-\tfrac{3}{2}G_2'(t)\right) = 0,
  \label{Liesym-det-eq-simplified2}
  \\
&& G_1'(y)-G_2'(t) = 0,
  \;\;
  G_{4y}(y,t) = 0,
  \;\;
  G_3'(y) = 0.
  \label{Liesym-det-eq-simplified3}
\end{eqnarray}
Solving
(\ref{Liesym-det-eq-simplified1})-(\ref{Liesym-det-eq-simplified3})
then results in
\begin{equation}
\label{unknown-functions}
\!\! G_1(y) =\! m_1 + m_5 y, \;
G_2(t) =\! m_2 + m_5 t, \;
G_3(y) =\! \tfrac{1}{2} m_5, \;
G_4(y,t) =\! m_3 +\! m_4 t,
\end{equation}
where the $m_i \; (i = 1, \cdots, 5)$
are arbitrary constants. The final expressions for
$\xi^1$, $\xi^2$, $\eta^1$ and $\eta^2$ then are
\begin{equation}
\label{solve-det-eqn-natural}
   \xi^1 = m_1 + m_5 y, \;\;
    \xi^2 = m_2 + m_5 t, \;\;
  \eta^1 = 0, \;\;
  \eta^2 = m_3 + m_4 t + m_5 u.
\end{equation}
Hence, the following Lie symmetries are obtained for
(\ref{magan-(3.7)-(3.8)}) for the case when $n>0 $ is any natural
number:
\begin{equation}
\label{Lie-sym}
\!\!\!X_1 = \frac{\partial }{\partial y},
\;\;
X_2 = \frac{\partial }{\partial t}, \;\;
X_3 = \frac{\partial }{\partial u}, \;\;
X_4 = t\frac{\partial}{\partial u}, \;\;
X_5 = y\frac{\partial }{\partial y}+ t\frac{\partial }{\partial t}
      + u\frac{\partial }{\partial u}.
\end{equation}
The generators $X_1$ and $X_2$ represent translations in time $y$
and space $t$, respectively. The displacement $u(y,t)$ appears
linearly in the model, and thus $X_3$ corresponds to a translation
in the variable $u$. $X_4$ corresponds to time-dependent changes in
displacement under the transformation $(y,t,\sigma, u) \rightarrow
(y,t, \sigma, u + t),$ where $a$ is an arbitrary constant. The
generator $X_5$ corresponds to a scaling symmetry and expresses the
scaling homogeneity of (\ref{magan-(3.7)-(3.8)}) under the
transformation $(y,t,\sigma, u) \rightarrow
(\frac{y}{\kappa},\frac{t}{\kappa}, \sigma, \kappa u)$. Note that
$\sigma$ does not occur in any of the generators in (\ref{Lie-sym}).
\vspace*{-2mm}
\section{Optimal system of one-dimensional subalgebras of the symmetry algebra for system  (\ref{magan-(3.7)-(3.8)}) }
\label{optical-system}
\vspace*{-2mm}
In this section, we derive the optimal system
of one-dimensional subalgebras for the symmetry algebra of the
system (\ref{magan-(3.7)-(3.8)}). This is a structured approach to
systematically reduce the original system of PDEs to simpler, often
solvable equations
\cite{ovsiannikov-academic-1982,olver-springer-1993,
patera-winternitz-jmp-1977}.

The Lie algebra for system (\ref{magan-(3.7)-(3.8)}) is five
dimensional. The Lie bracket/commutation relation for symmetry
generators $X_i$ and $X_j$ is defined as
\begin{equation}
[X_i, X_j] = X_i X_j - X_j X_i.
\label{lie_bracket}
\end{equation}
The commutation relations for the five-dimensional Lie algebra of
system (\ref{magan-(3.7)-(3.8)}) are given in
Table~\ref{Commutator_Table}, where the $(i,j)$-entry represents
$[X_i, X_j].$
\vskip 0.0001pt
\noindent
\vspace{-2mm}
\begin{table}[h!]
\centering
\renewcommand{\arraystretch}{1.5}
\setlength{\tabcolsep}{22.5pt}
\begin{tabular}{cccccc}
\hline
$[\phantom{x},\phantom{x}]$ & $X_1$ & $X_2$ & $X_3$ & $X_4$ & $X_5$ \\
\hline
$X_1$ & $0$   & $0$   & $0$   & $0$   & $X_1$ \\
$X_2$ & $0$   & $0$   & $0$   & $X_3$ & $X_2$ \\
$X_3$ & $0$   & $0$   & $0$   & $0$   & $X_3$ \\
$X_4$ & $0$   & $-X_3$& $0$   & $0$   & $0$ \\
$X_5$ & $-X_1$& $-X_2$& $-X_3$& $0$   & $0$ \\
\hline
\end{tabular}
\vspace{0.5cm} \caption{ Commutation relations for five-dimensional
Lie algebra of system (\ref{magan-(3.7)-(3.8)}).} \vspace{0.3cm}
\label{Commutator_Table}
\end{table}
\vskip 0.01pt
\noindent
The adjoint representation is computed using the commutation
relations from Table~\ref{Commutator_Table} and the familiar Lie
series (see, e.g.,
\cite{ovsiannikov-academic-1982,olver-springer-1993}):
\begin{equation}
{\mathrm{Ad}}(\exp(\epsilon X))Y
= Y-\epsilon[X,Y]+\tfrac{1}{2!}\epsilon^2[X,[X,Y]]
-\tfrac{1}{3!} \epsilon^3[X,[X,[X,Y]]] +....
\label{1m}
\end{equation}
For example,
\begin{eqnarray}
{\mathrm{Ad}}(\exp(\epsilon X_1))X_5 &=& X_5-\epsilon[X_1,X_5] +
\tfrac{1}{2!}\epsilon^2[X_1,[X_1,X_5]] -...
\nonumber \\
&=& X_5-\epsilon X_1, \label{2m}
\end{eqnarray}
which is shown in the first row and fifth column in
Table~\ref{adjoint_Table}.
Similarly, calculating the remaining entries of the adjoint table is
straightforward.
\vskip 0.01pt
\noindent
\begin{table}[hb]
\renewcommand{\arraystretch}{1.5}
\setlength{\tabcolsep}{12.5pt}
\begin{tabular}{cccccc}
\hline
Ad & $X_1$ & $X_2$ & $X_3$ & $X_4$ & $X_5$ \\
\hline
$X_1$ & $X_1$   & $X_2$   & $X_3$   & $X_4$  & $X_5-\epsilon X_1$ \\
$X_2$ & $X_1$   & $X_2$   & $X_3$   & $X_4-\epsilon X_3$   & $X_5-\epsilon X_2$ \\
$X_3$ &$X_1$   & $X_2$   & $X_3$   & $X_4$   & $X_5-\epsilon X_3$ \\
$X_4$ &$X_1$   & $X_2+\epsilon X_3$   & $X_3$   & $X_4$   & $X_5$ \\
$X_5$ & $X_1 \exp(\epsilon)$   & $X_2\exp(\epsilon)$ & $X_3\exp(\epsilon)$ & $X_4$   & $X_5$ \\
\hline
\end{tabular}
\vspace{0.5cm} \caption{ Adjoint representation of Lie algebra for
system (\ref{magan-(3.7)-(3.8)}). } \label{adjoint_Table}
\end{table}
\vskip 0.01pt
\noindent
Given a non-zero vector $X$
\begin{equation}
\label{generalized-sym}
X = a_1 X_1 + a_2 X_2 + a_3 X_3 + a_4 X_4 + a_5 X_5,
\end{equation}
where $a_i \; (i = 1,\cdots,5)$ are arbitrary constants.
Our goal is to simplify $X$ by annulling and setting
the coefficients $a_i$ to one wherever possible,
using adjoint mappings.
If we act on such a $X$ by ${\mathrm{Ad}}(\exp(\epsilon X_1))$ by
using the adjoint representation given in Table~\ref{adjoint_Table},
we can make the coefficient of $X_1$ vanish.
The action of ${\mathrm{Ad}}(\exp(\epsilon X_1))$ on $X$ yields
\begin{equation}
{\mathrm{Ad}}(\exp(\epsilon X_1))X
= (a_1-\epsilon a_5) X_1 + a_2 X_2 + a_3 X_3 + a_4 X_4 + a_5 X_5,
\label{Ad_action_X1}
\end{equation}
and coefficient of $X_1$ vanishes provided $a_1-\epsilon a_5=0$
which gives $\epsilon=\frac{a_1}{a_5}$ when $a_5\not=0$.
Using scaling, we can set $a_5=1$.
The vector $X$ in (\ref{generalized-sym}) simplifies into
\begin{equation}
X'={\tilde{a_2}}X_2+{\tilde{a_3}} X_3+{\tilde{a_4}}X_4+X_5
\label{simplified-sym1},
\end{equation}
for certain scalars ${\tilde{a_2}}$, ${\tilde{a_3}}$ and
${\tilde{a_4}}$ depending on $a_2$, $a_3$ and $a_4$.

The action of ${\mathrm{Ad}}(\exp(\epsilon X_4))$ on $X'$
yields
\begin{equation}
{\mathrm{Ad}}(\exp(\epsilon X_4))X' =
{\tilde{a_2}}X_2+({\tilde{a_3}}+\epsilon {\tilde{a_2}})
X_3+{\tilde{a_4}}X_4+X_5 \label{Ad_action_X4},
\end{equation}
and the coefficient of $X_3$
vanishes if ${\tilde{a_3}}+\epsilon {\tilde{a_2}}=0$.
So, $\epsilon=-\frac{{\tilde{a_3}}}{{\tilde{a_2}}}$
provided ${\tilde{a_2}}\not=0$.
The vector $X'$ in (\ref{simplified-sym1}) simplifies into
\begin{equation}
X'' = {\tilde{a_2}} X_2 + {\tilde{a_4}} X_4 + X_5.
\label{simplified-sym2}
\end{equation}
The action of ${\mathrm{Ad}}(\exp(\epsilon X_5))$ on $X''$ leads to
\begin{equation}
{\mathrm{Ad}}(\exp(\epsilon X_5))X''
= {\tilde{a_2}}\exp(\epsilon) X_2 + {\tilde{a_4}} X_4 + X_5,
\label{Ad_action_X5}
\end{equation}
and now depending on the sign of ${\tilde{a_2}}$, we can make the
coefficient of $X_2$ either $+1$ or $-1$.  Thus, any one-dimensional
subalgebra spanned by $X$ with $a_5\not=0$, ${\tilde{a_2}}\not=0$ is
equivalent to one spanned by $\pm X_2+{\tilde{a_4}}X_4+X_5$. For the
case where $a_5\not=0$ and ${\tilde{a_2}}=0$, the action of
${\mathrm{Ad}}(\exp(\epsilon X_3))$ on $X'$ in
(\ref{simplified-sym1}) makes the coefficient of $X_3$ in
\begin{equation}
{\mathrm{Ad}}(\exp(\epsilon X_3))X'
= ({\tilde{a_3}}- \epsilon)X_3+{\tilde{a_4}}X_4+X_5
\label{Ad_action_X3a}
\end{equation}
vanish provided that $\epsilon={\tilde{a_3}}$.

Thus, we arrive at a one-dimensional subalgebra spanned by
${\tilde{a_4}}X_4+X_5$. In other words, every one-dimensional
subalgebra generated by $X$ with $a_5\not=0$ is equivalent to the
subalgebra spanned by one-dimensional subalgebra
$\pm X_2+{\tilde{a_4}}X_4+X_5$ and ${\tilde{a_4}}X_4+X_5$.
This completes
the construction of one-dimensional subalgebras for $a_5\not=0$.
One can follow a similar procedure to obtain all one-dimensional
subalgebras for the $a_5=0$ case. After straightforward
calculations, the optimal system of one-dimensional subalgebras
are spanned by
\begin{eqnarray}
Y_1 &=& \lambda X_4+X_5 \pm X_2, \;\; Y_2 = X_5 + \lambda X_4,
\nonumber \\
Y_3 &=& \mu X_1  + X_4 \pm X_2, \;\; Y_4 =  X_4 \pm X_1,
\\
Y_5 &=& X_4, \;\;
Y_6 =\mu X_1 \pm X_2, \;\;
Y_7 = \mu X_1 \pm X_3,\;\;
Y_8 = X_1,
\nonumber
\end{eqnarray}
where ${\tilde{a_1}}=\mu$ and ${\tilde{a_4}}=\lambda$.  We use the
discrete symmetries of (\ref{magan-(3.7)-(3.8)}) to replace the $\pm
1$ in the optimal system by 1. Note that $\sigma$ does not appear in
any of the $Y_i$. Consequently, switching the sign of $\sigma$
(below) has no effect on the $Y_i$. In detail: $(y, t, \sigma, u)
\rightarrow (y, -t, -\sigma, -u)$ allows one to replace the $\pm 1$
by 1 in $Y_1$ and $Y_3$. Likewise, $(y, t, \sigma, u) \rightarrow
(-y, -t, \sigma, -u)$ does the same in $Y_4$. For $Y_6$ one can use
$(y, t, \sigma, u) \rightarrow (y, -t, \sigma, u)$. The $\pm 1$ in
$Y_7$ can be replaced by 1 using the symmetry $(y, t, \sigma, u)
\rightarrow (y, t, -\sigma, -u)$. Consequently, the optimal system
of one-dimensional subalgebras can be expressed as
\begin{eqnarray}
\!\!\!
Y_1 &=&
 \frac{\partial }{\partial t}+\lambda t\frac{\partial}{\partial u}
+ y\frac{\partial }{\partial y}+ t\frac{\partial}{\partial t}
+ u\frac{\partial }{\partial u},
\nonumber \\
\!\!\!
Y_2 &=&
 y\frac{\partial }{\partial y}+ t\frac{\partial }{\partial t}
 + u\frac{\partial }{\partial u} + \lambda t\frac{\partial }{\partial u},
\;\;
Y_3 = \mu \frac{\partial }{\partial y} + \frac{\partial}{\partial t}
      + t\frac{\partial }{\partial u},
\\
Y_4 &=&
 \frac{\partial}{\partial y} +t\frac{\partial }{\partial u},
\;\;
Y_5 =\! t\frac{\partial }{\partial u},
\;\;
Y_6 = \mu \frac{\partial}{\partial y} +\frac{\partial }{\partial t},
\;\;
Y_7 = \mu \frac{\partial}{\partial y}+ \frac{\partial }{\partial u},
\;\;
Y_8 = \frac{\partial }{\partial y}.
\nonumber
\end{eqnarray}
\vspace*{-4mm}
\section{Closed-form solutions of system (\ref{magan-(3.7)-(3.8)})
via the optimal system of one-dimensional subalgebras}
\label{closed-form-sols-Y2}
\vspace*{-2mm}
In this section, the closed-form solutions of system
(\ref{magan-(3.7)-(3.8)}) are established based on the optimal system
of one-dimensional subalgebras.
The calculations for $Y_1$ and $Y_6$ will be presented in detail,
while the results for the remaining $Y_i$ will be summarized in a
table to save space.

The invariant surface conditions for the generator $Y_1$ yield
\begin{equation}
\label{inv-gen-symm} y \sigma_y + (t + 1) \sigma_t =0, \;\; y u_y +
(t + 1) u_t = \lambda t + u .
\end{equation}
We have the following form of a group invariant solution:
\begin{equation}
\label{sigma-u-genX}
\sigma(y, t) = P_1 \left( \psi_1 \right), \;\;
u(y, t) = y Q_1 \left( \psi_1 \right)
  + \lambda\left( (t+ 1)\ln(y)+ 1\right),
\end{equation}
where $\psi_1$ is a similarity variable defined as $\psi_1 =
\frac{t+ 1}{y}$.

Substitution of \eqref{sigma-u-genX} into
(\ref{magan-(3.7)-(3.8)}) results in the following system of ODEs:
\begin{eqnarray}
&& \delta \psi_1 Q_1'-\delta Q_1-\delta \lambda \psi_1
+ P_1 \left(1 + P_1^2\right)^n =0,
\label{reduce-ode1}
\\
&& \delta Q_1'' +\psi_1 P_1' =0.
\label{reduce-ode2}
\end{eqnarray}
Differentiating \eqref{reduce-ode1} with respect to $\psi_1$, yields
\begin{equation}
\delta \psi_1 Q_1'' - \left(\delta \lambda \psi_1
- P_1 \left(1 + P_1^2 \right)^n \right)'=0.
\label{reduce-ode1a}
\end{equation}
Substituting $Q_1''$ from \eqref{reduce-ode1a} into
\eqref{reduce-ode2}, we get
\begin{equation}
 \left(\delta \lambda \psi_1
 - P_1 \left(1 + P_1^2\right)^n \right)'+\psi_1^2 P_1' = 0,
\label{reduce-ode2a}
\end{equation}
where prime denotes differentiation with respect to $\psi_1$.
The solution of ODE \eqref{reduce-ode2a} has the unknown function
$P_1(\psi_1)$. Next, we insert $P_1(\psi_1)$ into
\eqref{reduce-ode1} to find $Q_1(\psi_1)$.
For $\lambda \neq 0$, the reduction of
(\ref{magan-(3.7)-(3.8)}) to ODEs is possible;
however, to successfully derive closed-form solutions
we have to set $\lambda=0$.
\vspace*{-4mm}
\subsection{Result for arbitrary $n$}
\vspace*{-2mm}
When $\lambda=0$, \eqref{reduce-ode1} and \eqref{reduce-ode2a} for
arbitrary $n$ yield
\begin{equation}
\label{sol1-genX}
P_1(\psi_1) = C_1, \;\;
Q_1(\psi_1) = \tfrac{1}{\delta} C_1 (1 + C_1^2)^n + C_2 \psi_1,
\end{equation}
where $C_1$ and $C_2$ are arbitrary constants of integration.
Substituting $P_1(\psi_1)$ and $ Q_1(\psi_1)$ from \eqref{sol1-genX}
into \eqref{sigma-u-genX}, the final expressions for the group
invariant solutions for the variables $\sigma$ and $u$ (for
arbitrary $n$) are
\begin{equation}
\label{sigma-u-genX-Final-gen-n} \sigma(y, t) = C_1 , \;\; u(y, t) =
\tfrac{1}{\delta} C_1 (1 + C_1^2)^n y + C_2 (t+1).
\end{equation}
\vspace*{-6mm}
\subsection{Result for $n=1$}
\vspace*{-2mm}
For $\lambda=0$ and $n=1$, we obtain two additional solutions:
\begin{eqnarray}
\label{sol2-Y2-n1}
P_1(\psi_1) &=&
\pm \frac{1}{\sqrt{3}} \sqrt{\psi_1^2-1},
\\
Q_1(\psi_1) &=&
\left(C_3\mp \frac{\sqrt{3}}{12 \delta}\ln(3)\right)
\psi_1 \mp \frac{\sqrt{3}}{18
\delta}(\psi_1^2-4)\sqrt{\psi_1^2-1}
\nonumber \\
 &&\, \mp \frac{\sqrt{3}}{6 \delta}\psi_1\ln
\left(\sqrt{\psi_1^2-1}+\psi_1\right),
\end{eqnarray}
where  $C_3$ is an arbitrary constant of integration.
Substituting
$P_1(\psi_1)$ and $ Q_1(\psi_1)$ from (\ref{sol2-Y2-n1}) into
\eqref{sigma-u-genX}, the final expressions for the group invariant
solutions for
$\sigma$ and $u$ (for $n=1$ in
(\ref{magan-(3.7)-(3.8)})) are
\begin{eqnarray}
\!\!\!\! \sigma(y, t) &=& 
\pm \frac{\sqrt{(t+ 1)^2 - y^2}}{\sqrt{3} y},
\nonumber\\
\label{sigma-u-genY2-n1-u} \!\!\!\! u(y, t) &=& \left(C_3\mp
\frac{\sqrt{3}}{12\delta}\ln(3)\right)(t+ 1) \mp
\frac{\sqrt{3}}{18\delta} \frac{((t+1)^2 - 4 y^2) \sqrt{(t+1)^2 
- y^2}}{y^2 }
\nonumber \\
\!\!\!\! && \, \mp \frac{\sqrt{3}}{6\delta} (t+1)
\ln\left(\frac{\sqrt{(t+1)^2 - y^2} + t+1}{y}\right).
\end{eqnarray}
In the following section, we present closed-form solutions
corresponding to $Y_6$. They might hold physical significance
because they represent traveling wave solutions.
\vspace*{-2mm}
\section{Traveling wave solutions}
\label{traveling-waves} 
\vspace*{-2mm}
In this section, we derive the traveling
wave solutions of system (\ref{magan-(3.7)-(3.8)}) with $\beta=1$
using $Y_6$. We consider the variables
\begin{equation}
\label{sigma-u-case2}
\sigma(y, t) = P_6 \left(\psi_4\right),
\;\;
u(y,t) =  Q_6 \left(\psi_4\right),
\end{equation}
where $\psi_4 =y - \mu t$ is a similarity variable. Substitution of
equation \eqref{sigma-u-case2} into (\ref{magan-(3.7)-(3.8)}),
results in the following system of ODEs:
\begin{eqnarray}
&& \delta Q_6' - P_6 \left(1 + P_6^2\right)^n = 0,
\label{reduce-ode1-case2}
\\
&& \delta \mu^2 Q_6'' - P_6' = 0,
\label{reduce-ode2-case2}
\end{eqnarray}
where prime denotes differentiation with respect to $\psi_4$.
Differentiating
\eqref{reduce-ode1-case2} with respect to $\psi_4$ yields
\begin{equation}
\label{reduce-ode1a-case2}
\delta Q_6'' = \bigg( P_6 \left(1 + P_6^2\right)^n \bigg)'.
\end{equation}
Substituting \eqref{reduce-ode1a-case2} into
\eqref{reduce-ode2-case2}, yields
\begin{equation}
\label{reduce-ode2a-case2}
\mu^2 \bigg( P_6 \left(1 + P_6^2\right)^n \bigg)' - P_6'
= P_6' \bigg( \mu^2 \left(1 + P_6^2\right)^n
        + 2 n \mu^2 P_6^2 \left(1 + P_6^2\right)^{n-1} -1\bigg)
= 0.
\end{equation}
Hence, the cases $P_6' = 0$ and $P_6' \ne 0$ must be considered.
In the latter case, analytic solutions for $P_6$ can still be computed
for $n \leq 4$ but the complexity of the expressions drastically increases
as $n$ gets larger.
Therefore, we only report the solution for $n=1$ below.
\vskip 0.01pt
\noindent
\begin{table}[]
\centering
\begin{tabular}{llll}
\hline & Group invariant solution & Reduced ODEs
& Closed-form solution \\
\hline
$Y_1$ & \( \begin{array}{l}
         \sigma(y, t) \!=\! P_1 \left( \psi_1 \right)
         \\
         u(y, t) \!=\! y Q_1 \left( \psi_1 \right)
         \\
         \; + \lambda\left( (t\!+ \!1) \ln(y)\!+\!1\right)
         \\
         \psi_1 = \frac{t+ 1}{y}
         \\
\end{array} \) &
\( \begin{array}{l}
\\
(\ref{reduce-ode1})\!-\!(\ref{reduce-ode2})
\\ \\ \\
\end{array} \)
& \(\begin{array}{l} \text{solutions for} \; \lambda=0 \;
\text{only}
\\
(\ref{sigma-u-genX-Final-gen-n})
\\
(\ref{sigma-u-genY2-n1-u})
\\
\\
\end{array} \)
\\
\hline
$Y_2$ & \(\begin{array}{l}
\\
\sigma(y, t) \!=\! P_2 \left( \psi_2 \right)
\\
u(y, t) \!=\! t Q_2 \left( \psi_2 \right)
\\
\; + \lambda t \ln(t)
\\
\psi_2 = \frac{y}{t}
\\ \\
\end{array}\) & \(\begin{array}{l}
\\ \\
\delta Q_2'- P_2 \left(1 \!+\! P_2^2\right)^n \!=\! 0
\\
 \delta \psi_2^2 Q_2'' - P_2' +\delta \lambda  = 0
\\ \\ \\
\end{array}\) &
\(\begin{array}{l} \text{solutions for} \; \lambda=0 \; \text{only}
\\
\sigma(y,t) = C_4
\\
u(y,t) \!=\! \tfrac{1}{\delta} C_4 (1 \!+\! C_4^2)^n y \!+\! C_5 t
\\
\sigma(y,t) = \pm \frac{\sqrt{t^2 - y^2}}{\sqrt{3} y}
\\
u(y,t) = C_6 t
\\
\;\,\mp \frac{\sqrt{3}}{18\delta} \frac{(t^2 - 4 y^2)
\sqrt{t^2 - y^2}}{y^2 }
\\
\;\, \mp \frac{\sqrt{3}}{6\delta} t \tanh^{-1}\left(
\frac{t}{\sqrt{t^2 - y^2} }\right)
\end{array}\)
\\
\hline
$Y_3$ & \( \begin{array}{l} \sigma(y,t) = P_3(\psi_3)
\\
u(y,t)\!=\! \frac{t^2}{2} \!+\! Q_3(\psi_3)
\\
\psi_3 = y -\mu t
\end{array}\) &
\( \begin{array}{l} P_3' - \delta - \mu^2 \delta Q_3'' = 0
\\
\delta Q_3' \!-\! P_3 (1 \!+\! P_3^2)^n \!=\!0
\\
\end{array}\) &
\( \begin{array}{l}  \text{solutions for} \; \mu=0\; \text{only}\\
 \sigma(y,t) = \delta y+C_7
\\
u(y,t) = \frac{t^2}{2} \\
+\frac{1}{2\delta^2(n+1)} (1 + C_7^2+\delta^2 y^2\\
+ 2\delta C_7 y)^{n+1} + C_8
\end{array} \)
\\
\hline
$Y_4$ & \( \begin{array}{l} \sigma(y,t) =  P_4(t)
\\
u(y,t) =  t y + Q_4(t)
\end{array}\) &
\( \begin{array}{l}
 Q_4'' = 0
 \\
 t \delta - P_4 (1 \!+\! P_4^2)^n \!=\! 0
\end{array}\) &
\( \begin{array}{l} \sigma(y,t) = P_4(t)
\\
u(y,t) \!=\!  t y \!+\! C_9 t \!+\! C_{10}
\end{array} \)
\\
\hline
$Y_5$ & \phantom{xxxxxx} \text{---} & \phantom{xxxxxx} \text{---} & \phantom{xxxxxx} \text{---}
\\
\hline
$Y_6$ & \( \begin{array}{l} \sigma(y, t) = P_6\left(\psi_4\right)
\\
u(y,t) = Q_6\left(\psi_4\right)
\\
\psi_4 = y -\mu t
\end{array}\) &
\( \begin{array}{l}
(\ref{reduce-ode1-case2})\!-\!(\ref{reduce-ode2-case2})
\\ \\
\end{array}\) &
\( \begin{array}{l} (\ref{sigma-u-genX-Final-Case2})
\\
(\ref{sigma-u-genX-Final-Case2-n1})
\\
\end{array} \)
\\
\hline
$Y_7$ & \( \begin{array}{l} \sigma(y,t) =  P_7(t)
\\
u(y,t) =  \frac{ y}{\mu} + Q_7(t)
\end{array} \) &
\( \begin{array}{l}
 Q_7'' = 0
 \\
 \delta - \mu P_7 (1 + P_7^2)^n = 0
\end{array} \) &
\( \begin{array}{l} \sigma(y,t) =  P_7(t)
\\
u(y,t) \!=\!  \frac{ y}{\mu} \!+\! C_{14} t \!+\! C_{15}
\end{array} \)
\\
\hline
$Y_8$ & \( \begin{array}{l} \sigma(y,t) =  P_8(t)
\\
u(y,t) =  Q_8(t)
\end{array} \) &
\( \begin{array}{l}
 Q_8'' = 0 \\
 P_8 (1 + P_8^2)^n = 0
\end{array} \) &
\( \begin{array}{l}
\sigma(y,t) = 0 \\
u(y,t) = C_{16} t+ C_{17} \\
\end{array} \)
\\
\hline
\end{tabular}
\vspace{0.2cm} \caption{ Reductions and closed-form solutions of
(\ref{magan-(3.7)-(3.8)}) based on the optimal system of
one-dimensional subalgebras of the symmetry algebra. The first set
of solutions for $\sigma$ and $u$ (in the third column)  is for any
value of $n$; the second set is for $n=1$. }
\label{Solution-summary-Table}
\end{table}
\vspace*{-6mm}
\subsection{Result for arbitrary $n$}
\vspace*{-2mm}
Equations \eqref{reduce-ode1-case2} and \eqref{reduce-ode2a-case2}
(for arbitrary $n$) yield
\begin{equation}
\label{sol-genX-case2} P_6(\psi_4) = C_{11}, \;\; Q_6(\psi_4) =
\tfrac{1}{\delta} C_{11} (1 + C_{11}^2)^n \psi_4 + C_{12},
\end{equation}
where $C_{11}$ and $C_{12}$ are arbitrary constants of integration.
Substituting $P_6(\psi_4)$ and $ Q_6(\psi_4)$ from
\eqref{sol-genX-case2} into \eqref{sigma-u-case2},
the travelling wave solutions
for $\sigma$ and $u$ are
\begin{equation}
\label{sigma-u-genX-Final-Case2} \sigma(y, t) = C_{11}, \;\; u(y, t)
= \tfrac{1}{\delta} C_{11}(1 + C_{11}^2)^n (y-\mu t) + C_{12},
\end{equation}
where $\mu \neq 0$.
\vspace*{-2mm}
\subsection{Result for $n=1$}
\vspace*{-2mm}
For $n=1$, we have two additional solutions
\begin{equation}
\label{sol2-genX-case2-n1} P_6(\psi_4) = \pm
\frac{\sqrt{1-\mu^2}}{\sqrt{3}\mu}, \;\;
Q_6(\psi_4) = \pm \frac{1}{\delta}
\frac{(1+2\mu^2)}{3\sqrt{3}\mu^3} \sqrt{1-\mu^2} \, \psi_4 + C_{13},
\end{equation}
where $\mu \neq 0$ and $C_{13}$ are arbitrary constants of
integration. Substituting $P_6(\psi_4)$ and $Q_6(\psi_4)$ from
\eqref{sol2-genX-case2-n1} into \eqref{sigma-u-case2}, the
traveling solutions for $\sigma$ and $u$ are
\begin{equation}
\label{sigma-u-genX-Final-Case2-n1} \sigma(y, t) = \pm
\frac{\sqrt{1-\mu^2}}{\sqrt{3}\mu}, \;\; u(y, t) = \pm
\frac{1}{\delta} \frac{(1+2\mu^2)} {3\sqrt{3}\mu^3} \sqrt{1-\mu^2}\,
(y - \mu t) + C_{13},
\end{equation}
where $\mu\neq 0$.

A similar approach has been applied to the remaining generators $Y_i$.

For brevity, the results are summarized in
Table~\ref{Solution-summary-Table}, where the $C_i$ are arbitrary
constants of integration. Generator $Y_5$ does not provide any
reductions and group-invariant solutions.
\vspace*{-2mm}
\section{Computation of conservation laws using scaling homogeneity}
\label{cons-laws}
\vspace*{-2mm}
In this section we will show how to compute conservation laws for
(\ref{magan-(3.7)-(3.8)}) using the scaling symmetry
approach which originated in work by Kruskal and collaborators
\cite{gardner-etal-ii-jmp-1968,kruskal-etal-v-jmp-1970} and was
further developed by Hereman and co-workers (see, e.g.,
\cite{poole-hereman-jsc-2011,goktas-hereman-jsc-1997,hereman-etal-book-nova-2009}).
\vspace*{-2mm}
\subsection{Scaling homogeneity}
\label{scalinghomogeneity}
\vspace*{-2mm}
System (\ref{magan-(3.7)-(3.8)}) has a
two-parameter family of scaling (dilation) symmetries,
\begin{equation}
\label{magan-sys1-scale}
\!\!
(y, t, \sigma, u, \beta) \rightarrow
(\kappa^{-(2n+1)r+s} y, \kappa^{-(n+1)r+s} t,
\kappa^r \sigma, \kappa^s u, \kappa^{2r} \beta)
= ( {\tilde{y}}, {\tilde{t}}, {\tilde{\sigma}}, {\tilde{u}},
{\tilde{\beta}}),
\end{equation}
parameterized by the arbitrary real numbers $r$ and $s$.
The constant $\kappa \ne 0$ is an arbitrary scaling parameter.
Notice that if we had not introduced an auxiliary parameter
$\beta$ with an appropriate scale,
(\ref{magan-(3.7)-(3.8)})
would not be scaling homogeneous unless $r = 0$.
To verify that (\ref{magan-sys1-scale}) is correct, replace
$(y, t, \sigma, u, \beta)$ in terms of
$({\tilde{y}}, {\tilde{t}}, {\tilde{\sigma}}, {\tilde{u}}, {\tilde{\beta}})$,
yielding
\begin{equation}
\label{magan-sys1-scaling-test}
\kappa^{-2(n+1)r+s} \, \tilde{\sigma}_{\tilde{y}}
= \kappa^{-2(n+1)r+s} \,\delta \tilde{u}_{\tilde{t}\tilde{t}}, \quad
\kappa^{-(2n+1)r} \, \tilde{u}_{\tilde{y}}
= \kappa^{-(2n+1)r} \, \tfrac{1}{\delta} \tilde{\sigma}
(\tilde{\beta} + \tilde{\sigma}^2)^n,
\end{equation}
which, after cancellation of the common factors, is the same as
(\ref{magan-(3.7)-(3.8)}).

A quick way to compute (\ref{magan-sys1-scale}) is to introduce the
notions of weight, rank, and uniformity of rank.
The {\em weight}, $W$, of a variable is the exponent of $\kappa$
that multiplies that variable.
With regard to (\ref{magan-sys1-scale}), one has
$W(y) = -(2n+1)r+s $, $\, W(t) = -(n+1)r+s$, and
\begin{equation}
\label{magan-sys1-weights}
W(D_y) \!=\! (2n\!+\!1)r\!-\!s,
W(D_t) \!=\! (n\!+\!1)r\!-\!s,
W(\sigma) \!=\! r, W(u) \!=\! s,
W(\beta) \!=\! 2r.
\end{equation}
The {\em rank} of a monomial is defined as its total weight.
For example,
$(\beta + \sigma^2)^n$ has rank $2nr$.
A polynomial or equation is called {\em uniform in rank} if all
its monomials have equal ranks.

If the weights (\ref{magan-sys1-weights}) were not known yet,
they can be straightforwardly computed as follows:
Requiring that (\ref{magan-(3.7)-(3.8)}) is uniform in rank, yields
\begin{eqnarray}
\label{magan-sys1-weighteq1}
&& W(\sigma) + W(D_y) = W(u) + 2 W(D_t),
\\
\label{magan-sys1-weighteq2}
&& W(u) + W(D_y) = (2n+1) W(\sigma), \quad
W(\beta) = 2 W(\sigma).
\end{eqnarray}
Hence,
\begin{equation}
\label{magan-sys1-generalweights}
W(D_y) = (2n+1) W(\sigma) - W(u),
\;\;
W(D_t) = (n+1) W(\sigma) - W(u),
\end{equation}
where $W(\sigma)$ and $W(u)$ can be taken at liberty as long as
all weights in (\ref{magan-sys1-weighteq1})-(\ref{magan-sys1-weighteq2})
are strictly positive and, preferably, small integers.
The two-parameter family of scalings in (\ref{magan-sys1-scale})
arises by setting $W(\sigma) = r$ and $W(u) = s$.
Using weights to express uniformity in rank,
the scaling symmetry (\ref{magan-sys1-scale}) of
(\ref{magan-(3.7)-(3.8)}) can be computed using linear algebra.

To get the lowest possible weights, we take $r = 1$ and $s = n$ for
which (\ref{magan-sys1-weights}) simplifies into
\begin{equation}
\label{magan-sys1-weightssimple}
W(D_y) = n+1, \, W(D_t) = 1, \, W(\sigma) = 1,
\, W(u) = n, \, W(\beta) = 2.
\end{equation}
We will use both (\ref{magan-sys1-weights}) and (\ref{magan-sys1-weightssimple}) in the computations below.
\vspace*{-2mm}
\subsection{Conservation laws}
\label{conslaws}
\vspace*{-2mm}
A {\em conservation law} for (\ref{magan-(3.7)-(3.8)}) reads
\begin{equation}
\label{conslaw}
{\mathrm D}_{y} \, T^y + {\mathrm D}_{t} \, T^t \,\dot{=} \, 0,
\end{equation}
where ${\mathrm D}_{y}$ and ${\mathrm D}_{t}$ were defined in
(\ref{Dy-operator}) and (\ref{Dt-operator}). The notation
$\,\dot{=}\,$ means that the equality should only hold on solutions
$\sigma(y,t)$ and $u(y,t)$ of the system.
Since (\ref{magan-(3.7)-(3.8)}) is an evolution system in variable $y$,
we call $T^y$ a {\em conserved density} and $T^t$ the corresponding
{\em flux}. Both are functions of $\sigma$, and $u$, and their
partial derivatives with respect to $t$. Note that all
$y$-derivatives can be eliminated using the system. The density and
flux could also explicitly depend on $t$, e.g., $T^y = t \sigma$,
$\, T^t = \delta (u - t u_t)$, as discussed below.

Since (\ref{conslaw}) is {\em linear} in the densities (and fluxes)
a linear combination of densities with constant coefficients is also
a density, and vice versa. If a density has arbitrary coefficients
(e.g., powers of parameter $\beta$), it can be split into
independent densities. The algorithm discussed below produces
densities free of constant terms and without terms that could be
moved into the flux.  In what follows we show that, among others,
(\ref{magan-(3.7)-(3.8)}) has the following conservation laws
\begin{eqnarray}
\label{magan-sys1-conslaw1}
&& \!\!\!\!\!\!\!\!\!\! {\mathrm D}_{y} (\sigma)
+ {\mathrm D}_{t} (-\delta u_t ) \,\dot{=} \, 0,
\\
\label{magan-sys1-conslaw2}
&& \!\!\!\!\!\!\!\!\!\! {\mathrm D}_{y} (t \sigma)
+ {\mathrm D}_{t} (\delta (u - t u_t)) \,\dot{=} \, 0,
\\
\label{magan-sys1-conslaw3}
&& \!\!\!\!\!\!\!\!\!\!
{\mathrm D}_{y} (u \sigma_t)
+ {\mathrm D}_{t}
  \Big( -\tfrac{1}{2 (n+1) \delta} \left(
  (\beta + \sigma^2)^{n+1} - (n+1) \delta^2 (u_t^2 - 2 u u_{tt}) \right)
 \!\Big) \,\dot{=} \, 0,
\\
\label{magan-sys1-conslaw4}
&& \!\!\!\!\!\!\!\!\!\!
{\mathrm D}_{y}
  \left( (\beta \!+\! \sigma^2)^{n+1} \!-\! \beta^{n+1}
  \!+\! (n\!+\!1) \delta^2 u_t^2 \right)
  \!+\! {\mathrm D}_{t}
  \left(-2 (n\!+\!1) \delta \sigma (\beta \!+\! \sigma^2)^n u_t \right)
\,\dot{=} \, 0,
\\
\label{magan-sys1-conslaw5}
&& \!\!\!\!\!\!\!\!\!\!
{\mathrm D}_{y}
  \left(
  \sigma u (\beta + \sigma^2)^n  \sigma_t - \tfrac{1}{6} \delta^2 u_t^3
  \right)
\nonumber \\
&& \!\!\!\!\!\!\!\!\!\!
+ {\mathrm D}_{t}
  \left( - \tfrac{1}{3 \delta} \beta^{2n} \sigma^3 f(t;y)
 + \tfrac{1}{2} \delta \sigma (\beta + \sigma^2)^n (u_t^2 - 2 u u_{tt})
 \right)
 \,\dot{=} \, 0,
\end{eqnarray}
where
$f(t;y) = {}_2F_1(\tfrac{3}{2}, -2 n; \tfrac{5}{2};-\tfrac{\sigma^2}{\beta})$
is the Gauss hypergeometric function.
Note that $y$ serves as a parameter in $f(t;y)$ which is a solution
of the first-order, non-homogenous ordinary differential equation,
 \begin{equation}
 \label{ode-for-2F1}
  \sigma f^{\prime} + 3 \sigma_t f
  = 3 \left( 1 + \tfrac{\sigma^2}{\beta} \right)^{2n} \sigma_t,
\end{equation}
where $f^{\prime} = \frac{d\,f(t;y)}{dt}$.

The first conservation law is the first equation of
(\ref{magan-(3.7)-(3.8)}) itself.
The second one arises after multiplication of that equation
by $t$ and integration by parts of $t u_{tt}$.
They can be computed with \verb|ConservationLawsMD.m| by setting
$n = 1, 2, 3, \dots$ which will return (\ref{magan-sys1-conslaw1})
and (\ref{magan-sys1-conslaw2}).
They can also be computed with the multiplier approach
\cite{olver-springer-1993,naz-mahomed-mason-amc-2008,anco-bluman-ejam-2002,steudel-zn-1962}
using the {\em Maple} code \verb|GeM|, which also has to be run for
specific values of $n$.
The goal is to compute
multipliers, $\Lambda_1$ and $\Lambda_2$, such that
\begin{equation}
\label{multipliers}
\Lambda_1 \, E^1 + \Lambda_2 \, E^2
\,\, \dot{=} \,\,
{\mathrm D}_{y} \, T^y + {\mathrm D}_{t} \, T^t,
\end{equation}
with $E^1$ and $E^2$ in (\ref{alt-lie-magan-(3.7)-(3.8)}).
To do so, one must make an assumption about the arguments of the multipliers.
Then, compute and solve the determining PDEs for them. Next,
substitute $\Lambda_1$ and $\Lambda_2$ into (\ref{multipliers}) and,
finally, compute $T^y$ and $T^t$ (see
\cite{naz-mason-jnmp-2009,naz-mason-mahomed-narwa-10(5)-2009} for
worked examples).
For the first three conservation laws
we took $\Lambda_1(y,t,\sigma, u, \sigma_t, u_t)$ and the same
dependencies for $\Lambda_2$.
Using \verb|GeM|, one gets simple PDEs
for the multipliers which can be readily solved.
For (\ref{magan-sys1-conslaw1})
the multipliers are $\Lambda_1 = 1$ and $\Lambda_2 = 0$.
Conservation laws (\ref{magan-sys1-conslaw2}) and
(\ref{magan-sys1-conslaw3}) correspond to $\Lambda_1 = t$,
$\,\Lambda_2 = 0$, and $\Lambda_1 = -u_t $, $\, \Lambda_2 =
\sigma_t$, respectively.

Clearly, (\ref{magan-sys1-conslaw3})-(\ref{magan-sys1-conslaw5}) are
complicated because they depend on the exponent $n$. Deriving them
requires a computational strategy
\cite{poole-hereman-jsc-2011,goktas-hereman-jsc-1997,hereman-etal-book-nova-2009}
and the use of codes like \verb|InvariantsSymmetries.m|
\cite{goktas-hereman-code-invarsym-1997} or
\verb|ConservationLawsMD.m|
\cite{poole-hereman-code-conslawsmd-2009}.
In the latter two packages, a {\em scaling symmetry method} is implemented.
The key idea of the algorithm is that the densities and fluxes are
uniform in rank and so is the entire conservation law.
For example, using (\ref{magan-sys1-weightssimple}),
the density and flux in (\ref{magan-sys1-conslaw1})
have ranks $1$ and $n+1$, respectively,
and each of the two terms in the conservation law itself
has rank $n+2$.
For (\ref{magan-sys1-conslaw3}), these ranks are $n + 2, 2n + 2,$
and $2 n + 3$, respectively.

When working with \verb|ConservationLawsMD.m|, one only needs to give the ranks of the densities one wants to compute and specify whether or not they should explicitly depend on $t$ and $y$ (and, if applicable, also the highest degree
in $t$).
Using (\ref{magan-sys1-weightssimple}), the ranks of the corresponding
fluxes and the conservation laws then follow from
\begin{eqnarray}
\label{rank-flux}
&&
{\mathrm{rank}} \, T^t = n + {\mathrm{rank}} \, T^y,
\\
\label{rank-conslaw}
&& {\mathrm{rank}}\, ( {\mathrm D}_y T^y + {\mathrm D}_t T^t )
= n + 1 + {\mathrm{rank}} \, T^y
= 1  + {\mathrm{rank}} \, T^t,
\end{eqnarray}
and both can be used for verification purposes.

For (\ref{magan-sys1-conslaw2}) through (\ref{magan-sys1-conslaw5}),
using (\ref{magan-sys1-weightssimple}) the respective ranks
of the densities are $0$, $n+2$, $2n+2$, and $3n+3$.
That homogeneity in rank is due to the fact that the defining equation (\ref{conslaw})
must be evaluated on solutions of
(\ref{magan-(3.7)-(3.8)}).
Consequently, densities, fluxes, and conservation laws themselves inherit
(or adopt) the scaling homogeneity of that system (and all its other
continuous and discrete symmetries).

As far as we know, there is no symbolic code available to compute
conservation laws for (\ref{magan-(3.7)-(3.8)}) with undetermined
exponent $n$. Based on the conservation laws computed with
\verb|ConservationLawsMD.m| for $n = 1, 2,$ and $3$ (and larger
values, if needed), it is often straightforward to guess the density
for arbitrary $n$ and compute the matching flux. In some cases, it
is easier to recognize the expression of the flux for arbitrary $n$
and then compute the density.
\vspace*{-2mm}
\subsection{Strategy to compute (\ref{magan-sys1-conslaw4})}
\label{compute-conslaw4}
\vspace*{-2mm} 
We first show how to compute
(\ref{magan-sys1-conslaw4}) for $n=1$ with
\verb|ConservationLawsMD.m|. Note that for $n=1$ in
(\ref{magan-sys1-weightssimple}) the density in
(\ref{magan-sys1-conslaw4}) has rank $4$ and does not explicitly
depend on $t$. We provide this information together with $W(\sigma)
= W(u) = n = 1$, and specify that $\beta$ should be treated as a
parameter with weight.  Without further intervention of the user,
the computation for $n=1$ proceeds in four steps. Once the density
and flux for $n=1$ are computed, the process is repeated for $n=2$
and $3$ and in Step 5 the conservation law
(\ref{magan-sys1-conslaw4}) for arbitrary $n$ is obtained by pattern
matching.
 \vskip 3pt \noindent {\bf Step 1}: The code uses
(\ref{magan-sys1-weightssimple}) to construct candidate densities
$T^y$ as linear combinations of scaling homogenous monomials
involving $\beta, \sigma $, and $u$ and their $t$-derivatives so
that each monomial has rank $4$. Each candidate density is free of
trivial (constant) terms and monomials that are $t$-derivatives
because the latter can be moved into $T^t$. To have the shortest
possible densities, monomials that only differ by a $t$-derivative
are also removed.

In more detail, the code first creates a list of the 13 monomials, $
\{ \beta \sigma^2, \beta \sigma u $, $ \beta u^2, \sigma^4 $, $
\sigma^3 u, \sigma^2 u^2 $, $ \sigma u^3, u^4 $, $ \sigma_t^2,
\sigma u \sigma_t $, $ u^2 \sigma_t, \sigma_t u_t, u_t^2 \} $, of
rank 4. Next, using (\ref{magan-sys1-generalweights}) with {\em
arbitrary} $r$ and $s$, the code splits these monomials according to
their ranks. For example, the monomials $\{ \beta \sigma^2,
\sigma^4,$ $ \sigma u \sigma_t, u_t^2 \}$ have rank $4r$, leading to
candidate density
\begin{equation}
\label{Ty4n1-candidate}
T^y
= c_1 \beta \sigma^2 + c_2 \sigma^4 + c_3 \sigma u \sigma_t
+ c_4 u_t^2,
\end{equation}
where $c_1$ through $c_4$ are undetermined coefficients.
This is the only density among the seven possible candidates
with ranks
$4r$, $3r+s$, $2r+2s$, $r+3s$, $4s$, $6r-2s$, and $5r-s$, respectively,
that eventually leads to a (non-zero) conservation law of rank 4.
Therefore, our discussion continues with (\ref{Ty4n1-candidate}).
Complete details on how densities are constructed algorithmically
can be found in
\cite{hereman-goktas-mca-2024,poole-hereman-jsc-2011,goktas-hereman-jsc-1997}.
\vskip 3pt
\noindent
{\bf Step 2}:
To compute the undetermined coefficients, the code computes
\begin{equation}
\label{DyTy4n1}
{\mathrm D}_y T^y =
(2 \beta c_1 \sigma + 4 c_2 \sigma^3 + c_3 u \sigma_t) \sigma_y
+ c_3 \sigma \sigma_t u_{y}
+ c_3 \sigma u \sigma_{ty}
+ 2 c_4 u_t u_{ty}
\end{equation}
and, using (\ref{magan-(3.7)-(3.8)}),
replaces $\sigma_y$, $u_y$, $\sigma_{ty} = \sigma_{yt}$, and $u_{ty}$,
to get
\begin{eqnarray}
\label{DyTy4n1-on-system}
P &=& \tfrac{1}{\delta} c_3 \sigma^2 (\beta + \sigma^2) \sigma_t
+ \tfrac{2}{\delta} c_4 (\beta  + 3 \sigma^2) \sigma_t u_t
+ \delta (2 \beta c_1 \sigma + 4 c_2 \sigma^3 + c_3 u \sigma_t) u_{tt}
\nonumber \\
&& + \delta c_3 \sigma  u u_{ttt}.
\end{eqnarray}
Now, $P = {\mathrm D}_y T^y$ must match $-{\mathrm D}_t T^t$ for
some flux $T^t$ (computed in Step 3 below). Since $P$ must be {\em
exact}, i.e., a {\em total} $t$-derivative of some expression, the
Euler operator (variational derivative)
\cite{hereman-etal-book-nova-2009,hereman-etal-book-birkhauser-2005}
for each of the dependent variables\footnote{At this point, $y$ is a
parameter in the dependent variables. Suppressing $y$, we write
$\sigma(t)$ and $u(t)$. } applied to $P$ must be zero. The code
applies the Euler operator for $\sigma(t)$
\begin{eqnarray}
\label{euler-sigma}
{\cal{E}}_{\sigma(t)}
&=& \sum_{k=0}^{K}
  (-{\mathrm D}_t)^k \frac{\partial }{\partial \sigma_{kt} }
  \nonumber \\
&=& \frac{\partial }{\partial \sigma}
  - {\mathrm D}_t \frac{\partial }{\partial \sigma_t}
  + {\mathrm D}_{t}^2 \frac{\partial }{\partial \sigma_{tt}}
  - {\mathrm D}_{t}^3 \frac{\partial }{\partial \sigma_{ttt}}
+ \ldots,
\end{eqnarray}
to $P$ where $K=1$ is the order of $\sigma$ in $t$. Explicitly, for
$P$ in (\ref{DyTy4n1-on-system}),
\begin{eqnarray}
\label{eulersigmat-on-E}
{\cal{E}}_{\sigma(t)} P
&=& \frac{\partial P}{\partial \sigma}
  - {\mathrm D}_t \frac{\partial P}{\partial \sigma_t}
\nonumber \\
&=& \tfrac{2}{\delta} \beta ( \delta^2 c_1 - c_4) u_{tt} - \delta
c_3 u_t u_{tt} + \tfrac{6}{\delta} (2 \delta^2 c_2 - c_4) \sigma^2
u_{tt}.
\end{eqnarray}
Next, the Euler operator for $u(t)$ is applied to $P$ which has a
third-order term $u_{ttt}$.
Hence, $K=3$ and
\begin{eqnarray}
\label{euleru-on-E}
\!\!\!\!\!\!\!{\cal{E}}_{u(t)} P
&=&
\frac{\partial P}{\partial u}
  - {\mathrm D}_t \frac{\partial P}{\partial u_t}
  + {\mathrm D}_{t}^2 \frac{\partial P}{\partial u_{tt}}
  - {\mathrm D}_{t}^3 \frac{\partial P}{\partial u_{ttt}}
\nonumber \\
\!\!\!\!\!\!\!&=& \tfrac{2}{\delta} \beta (\delta^2 c_1\!-\!c_4)
\sigma_{tt} \!-\!\delta c_3 (\sigma_t u_{tt}\!+\! \sigma_{tt} u_t)
\!+\!\tfrac{6}{\delta} (2 \delta^2 c_2\!-\!c_4) \sigma
  (2 \sigma_t^2\!+\!\sigma \sigma_{tt}).
\end{eqnarray}
Both expressions must vanish identically on the jet space where all
monomials in $\sigma$, $u$, $\sigma_t$, $u_t$, $\sigma_{tt}$,
$u_{tt}$, etc., are treated as independent. Then,
${\cal{E}}_{\sigma(t)} P \equiv 0$ and ${\cal{E}}_{u(t)} P \equiv 0$
yield the {\em linear} system $ \delta^2 c_1 - c_4 = 0$, $\, c_3 =
0$, and $2 \delta^2 c_2 - c_4 = 0,$ where $c_4$ is arbitrary
(confirming that any scalar multiple of $T^y$ is still a density).
To avoid fractions, the code takes $c_4 = 2 \delta^2$, and
substitutes the solution $c_1 = 2, c_2 = 1, c_3 = 0$, and $c_4 = 2
\delta^2$ into (\ref{Ty4n1-candidate}), yielding
\begin{equation}
\label{Ty4n1-code}
T^y
 = 2 \beta \sigma^2 + \sigma^4 + 2 \delta^2 u_t^2
 = (\beta + \sigma^2)^2 - \beta^2 + 2 \delta^2 u_t^2,
\end{equation}
 which matches $T^y$ in (\ref{magan-sys1-conslaw4}) for
$n = 1$.

To prepare for the computation of the flux (in the next step), the
constants are also substituted into (\ref{DyTy4n1-on-system}),
yielding
\begin{equation}
\label{DyTy4n1-final}
P = 4 \delta \left(
  (\beta  + 3 \sigma^2) \sigma_t u_t
  + \sigma (\beta + \sigma^2) u_{tt}
\right).
\end{equation}
\vskip 2pt \noindent {\bf Step 3}: Since $P = {\mathrm D}_y T^y = -
{\mathrm D}_t T^t$, to compute the flux $T^t$ the code must
integrate (\ref{DyTy4n1-final}) with respect to $t$ and reverse the
sign. For this simple example, {\em Mathematica} does this
flawlessly and returns
\begin{equation}
\label{Tt4n1-final}
T^t = - 4 \delta \sigma (\beta + \sigma^2) u_t,
\end{equation}
which matches $T^t$ in (\ref{magan-sys1-conslaw4}) for $n = 1$.

For expressions more complicated than (\ref{DyTy4n1-final}), {\em
Mathematica} often fails this task. Hence,
\verb|ConservationLawsMD.m| does not relay on {\em Mathematica}'s
built-in (black-box) routines for integration by parts. Instead, it
uses a sophisticated way to reduce the integration with respect to
$t$ to a one-dimensional integral with respect to a scaling
parameter using the so-called {\em homotopy operator}. This is a
tool from differential geometry \cite[p.\ 372]{olver-springer-1993}
to carry out integration by parts on the jet space. It is usually
presented in the language of differential forms but can translated
in standard calculus and, as such, has been used effectively for the
computation of conservation laws (see,
\cite{hereman-goktas-mca-2024,poole-hereman-jsc-2011,hereman-etal-book-nova-2009,hereman-etal-book-birkhauser-2005,
hereman-etal-mcs-2007,poole-hereman-aa-2010}).

Application of the homotopy operator requires the computation of two
integrands (one for $\sigma$, the other for $u$), followed by a
simple scaling of the dependent variables (and their derivatives),
and finally, a one-dimensional integral with respect to a scaling
parameter $\lambda$ (not to be confused with $\kappa$ in
(\ref{magan-sys1-scale})). In terms of the homotopy operator,
\begin{equation}
\label{homotopysigmauE}
T^t = - {\cal H}_{{\bf u}(t)} P
  = - \int_{0}^{1} ( I_{\sigma(t)} P
  + I_{u(t)} P )[\lambda {\bf u}] \,\frac{d \lambda}{\lambda},
\end{equation}
where ${\bf u}(t) = (\sigma(t), u(t))$ and $[\lambda {\bf u}]$
means that in the integrands one must replace
$\sigma$ by $\lambda \sigma$, $u$ by $\lambda u$,
$\sigma_t$ by $\lambda \sigma_t $, $u_t$ by $\lambda u_t$, etc.

The integrand for $\sigma(t)$ is given
\cite{poole-hereman-jsc-2011,hereman-etal-book-nova-2009,hereman-etal-mcs-2007}
by
\begin{eqnarray}
\label{integrandhomotopysigmaE}
I_{\sigma(t)} P
&=& \sum_{k=1}^{K}
  \left( \sum_{i=0}^{k-1} \sigma_{it} (-{\mathrm D}_t)^{k-(i+1)} \right)
  \frac{\partial P}{\partial \sigma_{kt}}
\nonumber \\
&=&
 (\sigma {\mathrm I})(\frac{\partial P}{\partial \sigma_t})
  + (\sigma_t {\mathrm I}
  - \sigma {\mathrm D}_t) (\frac{\partial P}{\partial \sigma_{tt}})
  + \ldots,
\end{eqnarray}
where ${\mathrm I}$ is the identity operator.
For (\ref{DyTy4n1-final}) where $K=1$, the software readily computes
\begin{equation}
\label{integrandhomotopysigmaE-eval}
I_{\sigma(t)} P
= (\sigma {\mathrm I}) (\frac{\partial P}{\partial \sigma_t})
= 4 \delta \sigma (\beta + 3 \sigma^2) u_t.
\end{equation}
With a formula similar to (\ref{integrandhomotopysigmaE})
for $u(t)$ and $K=2$,
\begin{equation}
\label{integrandhomotopyuE-eval}
I_{u(t)} P
= (u {\mathrm I})(\frac{\partial P}{\partial u_t})
  + (u_t {\mathrm I} - u {\mathrm D}_t)
  (\frac{\partial P}{\partial u_{tt}})
= 4 \delta \sigma (\beta + \sigma^2) u_t.
\end{equation}
Finally, using (\ref{homotopysigmauE}),
\begin{eqnarray}
\label{homotopysigmauE-eval}
T^t &=&
 - 8 \delta \int_{0}^{1} \left( \sigma (\beta + 2 \sigma^2) u_t \right)
 [\lambda {\bf u}] \, \frac{d \lambda}{\lambda}
= -8 \delta \sigma u_t
\int_{0}^{1} (\beta \lambda + 2 \lambda^3 \sigma^2) \, d \lambda
\nonumber \\
&=& - 4 \delta \sigma (\beta + \sigma^2) u_t,
\end{eqnarray}
which is exactly the flux in (\ref{Tt4n1-final}). \vskip 5pt
\noindent {\bf Step 4}: Once the density (\ref{Ty4n1-code}) and flux
(\ref{homotopysigmauE-eval}) are computed the code verifies that
they indeed satisfy (\ref{conslaw}). \vskip 5pt \noindent 
{\bf Step 5}: To determine (\ref{magan-sys1-conslaw4}) for arbitrary $n$, 
it suffices to compute the conservation laws for $n=2$ and $n=3$ for
(\ref{magan-(3.7)-(3.8)}) and do some pattern matching.

For $n=2$, requesting a density of rank $6$,
\verb|ConservationLawsMD.m| returns
\begin{eqnarray}
\label{Tyn2rank6}
T^y &=& 3 \beta^2 \sigma^2 + 3 \beta \sigma^4
    + \sigma^6 + 3 \delta^2 u_t^2
    = (\beta + \sigma^2)^3 - \beta^3 + 3 \delta^2 u_t^2,
\\
\label{Ttn2rank8}
T^t &=& -6 \delta \sigma (\beta + \sigma^2)^2  u_t.
\end{eqnarray}
 For $n=3$, asking for a density of rank $8$, the code produces
\begin{eqnarray}
\label{Tyn3rank8}
T^y &=& 4 \beta^3 \sigma^2 + 6 \beta^2 \sigma^4 + 4 \beta \sigma^6
        + \sigma^8 + 4 \delta^2 u_t^2
     = (\beta + \sigma^2)^4 - \beta^4 + 4 \delta^2 u_t^2,
\\
\label{Ttn2rank11}
T^t &=& -8 \delta \sigma (\beta + \sigma^2)^3 u_t.
\end{eqnarray}
  Inspecting (\ref{Ty4n1-code}), (\ref{Tyn2rank6}), and
(\ref{Tyn3rank8}), the density for
arbitrary $n$ in (\ref{magan-sys1-conslaw4}) is easy
to recognize.
Likewise, from (\ref{Tt4n1-final}), (\ref{Ttn2rank8}), and (\ref{Ttn2rank11})
the flux in (\ref{magan-sys1-conslaw4}) becomes obvious.

Use of \verb|ConservationLawsMD.m| requires $n$ to be positive integer.
However, once (\ref{magan-sys1-conslaw4}) is established,
it is also valid for rational values of $n > 0$,
provided the conservation law can be validated on solutions of
(\ref{magan-(3.7)-(3.8)}).
Testing for $n = \tfrac{1}{2}$ and $n = \tfrac{1}{3}$ confirmed that
(\ref{magan-sys1-conslaw4}) is indeed valid.
Therefore, our results apply to model equations involving, e.g.,
square roots and cubic roots.
\vspace*{-2mm}
\subsection{Computation of conservation law (\ref{magan-sys1-conslaw5})}
\label{compute-conslaw5}
\vspace*{-2mm}
Using \verb|ConservationLawsMD.m| for 
$n = 1, 2$, and $3$, a systematic search for conserved densities of 
high rank, generated the following results:
\begin{eqnarray}
\label{Tyn1rank6}
T^y &=& \sigma u (\beta + \sigma^2) \sigma_t
- \tfrac{1}{6} \delta^2 u_t^3,
\\
\label{Ttn1rank7}
T^t &=& -\tfrac{1}{\delta} \left(
       \tfrac{1}{3} \beta^2 \sigma^3
       + \tfrac{2}{5} \beta \sigma^5
       + \tfrac{1}{7} \sigma^7
       \right)
 + \tfrac{1}{2} \delta \sigma (\beta + \sigma^2)
 (u_t^2 - 2 u u_{tt}),
\end{eqnarray}
for $n=1$ when searching for a density of rank $6$;
\begin{eqnarray}
\label{Tyn1rank9}
T^y &=& \sigma u (\beta + \sigma^2)^2 \sigma_t
     - \tfrac{1}{6} \delta^2 u_t^3,
\\
\label{Ttn1rank11}
T^t &=& - \tfrac{1}{\delta}
        \left(
        \tfrac{1}{3} \beta^4 \sigma^3
        + \tfrac{4}{5} \beta^3 \sigma^5
        + \tfrac{6}{7} \beta^2 \sigma^7
        + \tfrac{4}{9} \beta \sigma^9
        + \tfrac{1}{11} \sigma^{11}
        \right)
\nonumber \\
&& + \tfrac{1}{2} \delta \sigma (\beta + \sigma^2)^2
   (u_t^2 - 2 u u_{tt}),
\end{eqnarray}
for $n=2$ and a density of rank $9$; and
\begin{eqnarray}
\label{Tyn1rank12}
\!\!\!\!\!\!\!
  T^y &=&
  \sigma u (\beta + \sigma^2)^3  \sigma_t -\tfrac{1}{6} \delta^2 u_t^3,
\\
\label{Ttn1rank15}
\!\!\!\!\!\!\!
T^t &=&
  - \tfrac{1}{\delta} \left(
  \tfrac{1}{3} \beta^6 \sigma^3
  + \tfrac{6}{5} \beta^5 \sigma^5
  + \tfrac{15}{7} \beta^4 \sigma^7
  + \tfrac{20}{9} \beta^3 \sigma^9
  + \tfrac{15}{11} \beta^2 \sigma^{11}
  + \tfrac{6}{13} \beta \sigma^{13}
  + \tfrac{1}{15} \sigma^{15}
  \right)
\nonumber \\
\!\!\!\!\!\!\!
&&
 + \tfrac{1}{2} \delta \sigma (\beta + \sigma^2)^3
 (u_t^2 - 2 u u_{tt}),
\end{eqnarray}
for $n=3$ and a density of rank $12$.

Inspecting (\ref{Tyn1rank6}), (\ref{Tyn1rank9}), and (\ref{Tyn1rank12}),
the form
\begin{equation}
\label{Tyn-rank3nplus3}
T^y = \sigma u (\beta + \sigma^2)^n \sigma_t
- \tfrac{1}{6} \delta^2 u_t^3,
\end{equation}
for arbitrary $n$ is obvious. Comparing  (\ref{Ttn1rank7}),
(\ref{Ttn1rank11}), and (\ref{Ttn1rank15}), the pattern of the last
term is equally clear but the first term requires further
investigation. Noticing the common factor $\sigma^3$ and leading
coefficient $-\tfrac{1}{3\delta} \beta^{2n}$, we assume
\begin{equation}
\label{Ttn-rank4nplus3-form1}
T^t = - \tfrac{1}{3 \delta} \beta^{2n} \sigma^3 F(y,t)
 + \tfrac{1}{2} \delta \sigma (\beta + \sigma^2)^n
 (u_t^2 - 2 u u_{tt}),
\end{equation}
and compute the equation for the unknown function $F(y,t).$
Substituting (\ref{Tyn-rank3nplus3}) and
(\ref{Ttn-rank4nplus3-form1}) into (\ref{conslaw})
yields
\begin{equation}
\label{ode-for-F}
 \sigma F_t + 3 \sigma_t F
  = 3 \left( 1 + \tfrac{\sigma^2}{\beta} \right)^{2n} \sigma_t.
\end{equation}
which is an ODE for $F(y,t) \equiv f(t;y)$ which matches (\ref{ode-for-2F1}).
Asking {\em Mathematica} to solve the ODE yields
\begin{equation}
\label{conslaw5-2F1}
F(y,t) = {}_2F_1(\tfrac{3}{2}, -2 n; \tfrac{5}{2};-\tfrac{\sigma^2}{\beta})
+ \tfrac{c(y)}{\sigma^3},
\end{equation}
where $c(y)$ is an arbitrary integration constant which we set to
zero to avoid a constant in $T^t$ in (\ref{Ttn-rank4nplus3-form1}).
Hence, (\ref{conslaw5-2F1}) confirms the result in
(\ref{magan-sys1-conslaw5}).

Alternatively, $T^t$ can be computed as follows.
Substituting
\begin{equation}
\label{Ttn-rank4nplus3-form2}
T^t = G(y,t) + \tfrac{1}{2} \delta \sigma
(\beta + \sigma^2)^n (u_t^2 -2 u u_{tt}),
\end{equation}
into (\ref{conslaw}) requires
\begin{equation}
\label{eq-for-G}
G_t + \tfrac{1}{\delta} (\beta + \sigma^2)^{2n} \sigma^2 \sigma_t = 0.
\end{equation}
Hence,
\begin{equation}
\label{integral-for-G}
G = - \tfrac{1}{\delta}
  \int \sigma^2 (\beta + \sigma^2)^{2n} \sigma_t \,dt
  = - \tfrac{1}{3 \delta} \beta^{2n} \sigma^3 \,
    {}_2F_1(\tfrac{3}{2}, -2 n; \tfrac{5}{2}; -\tfrac{\sigma^2}{\beta}),
\end{equation}
after setting the integration constant equal to zero.
Substitution of $G$ into (\ref{Ttn-rank4nplus3-form2}) then yields
the flux in (\ref{magan-sys1-conslaw5}).

Yet another way to compute the flux is to substitute
(\ref{Tyn-rank3nplus3}) and
\begin{equation}
\label{Ttn-rank4nplus3-form3}
T^t = H(\sigma) + \tfrac{1}{2} \delta \sigma
(\beta + \sigma^2)^n (u_t^2 -2 u u_{tt}),
\end{equation}
with unknown $H(\sigma)$ into (\ref{conslaw}) yielding
\begin{equation}
\label{eqforH}
H^{\prime}
= - \tfrac{1}{\delta} \left( \sigma (\beta + \sigma^2)^n \right)^2.
\end{equation}
Integration gives
\begin{equation}
\label{solHsigma}
H(\sigma) =
- \tfrac{1}{\delta} \int \left( \sigma (\beta + \sigma^2)^n \right)^2
\, d \sigma
= - \tfrac{1}{3 \delta} \beta^{2n} \sigma^3 \,
  {}_2F_1(\tfrac{3}{2}, -2 n; \tfrac{5}{2}; -\tfrac{\sigma^2}{\beta})
  + c,
\end{equation}
where the integration constant $c$ can be set to zero to avoid a constant
term in flux (\ref{magan-sys1-conslaw5}).
Substitution of $H$ into (\ref{Ttn-rank4nplus3-form3}) yields the
flux in (\ref{magan-sys1-conslaw5}).

For $n=\tfrac{1}{2}$ and $n=\tfrac{1}{4}$, (\ref{magan-sys1-conslaw5})
simplifies into
\begin{eqnarray}
\label{conslaw5-nonehalf}
&& {\mathrm D}_{y}
  \left(
  \sigma u \, \sqrt{\beta + \sigma^2}\, \sigma_t
   -\tfrac{1}{6} \delta^2 u_t^3
  \right)
  + {\mathrm D}_{t}
  \left(
  - \tfrac{1}{\delta}
  (\tfrac{1}{3} \beta \sigma^3 + \tfrac{1}{5} \sigma^5)
  \right.
\nonumber \\
&& \left.
  + \tfrac{1}{2} \delta \sigma \, \sqrt{\beta + \sigma^2}\,
  (u_t^2 - 2 u u_{tt}) \right) \,\dot{=} \, 0,
\end{eqnarray}
and
\begin{eqnarray}
\label{conslaw5-nonquarter}
&& {\mathrm D}_{y}
  \left(
  \sigma u \, \sqrt[4]{\beta + \sigma^2} \, \sigma_t
   - \tfrac{1}{6} \delta^2 u_t^3
  \right)
  + {\mathrm D}_{t}
  \left(
  \tfrac{1}{8 \delta}
  \left(
  \beta^2 {\mathrm{sinh}^{-1}}(\tfrac{\sigma}{\sqrt{\beta}})
  \right.
  \right.
\nonumber \\
&&
 \left.
 \left.
  - \sigma (\beta + 2 \sigma^2) \,\sqrt{\beta + \sigma^2}
  \right)
 + \tfrac{1}{2} \delta \sigma \,\sqrt[4]{\beta + \sigma^2} \,
 (u_t^2 - 2 u u_{tt})
 \right) \,\dot{=} \, 0.
\end{eqnarray}
Notice that for fractional values of $n$, the conservation laws
are no longer polynomial and that the last one involves the
inverse of hyperbolic function.
For $n=\tfrac{1}{3}$, {\em Mathematica} replaces
$f(t;y) = {}_2F_1(\tfrac{3}{2}, -\tfrac{2}{3}; \tfrac{5}{2}; -\tfrac{\sigma^2}{\beta})$ in (\ref{magan-sys1-conslaw5}) by
${}_2F_1(-\tfrac{2}{3}, \tfrac{3}{2};
\tfrac{5}{2}; -\tfrac{\sigma^2}{\beta})$ but does not further
simplify that hypergeometric function.
\vspace*{-2mm}
\subsection{Additional conservation laws}
\label{compute-newconslaw}
\vspace*{-2mm} 
In the section we present two additional
conservation laws of (\ref{magan-(3.7)-(3.8)}) for arbitrary $n$.

Using a variant of the strategy in Section~\ref{compute-conslaw5},
it is possible to find a density of rank $2n+3$ and the matching flux.

To do so, use \verb|ConservationLawsMD.m| to compute density-flux pairs
for $n=1, 2,$ and $3$, yielding
\begin{eqnarray}
\label{Tyn1rank5}
T^y &=& \sigma^3 (\tfrac{1}{3} \beta + \tfrac{3}{10} \sigma^2)
        + \delta \sigma u_t^2,
\\
\label{Ttn1rank6}
T^t &=& -\delta \sigma^2 (\beta + \tfrac{3}{2} \sigma^2) u_t
        -\tfrac{1}{3} \delta^3 u_t^3
\nonumber \\
    &=& \tfrac{1}{2} \delta \left(
       (\beta - 3 \sigma^2) (\beta + \sigma^2) - \beta^2 \right) u_t
       - \tfrac{1}{3} \delta^3 u_t^3,
\end{eqnarray}
for $n=1$ when searching for a density of rank $5$;
\begin{eqnarray}
\label{Tyn1rank7}
T^y &=& \sigma^3 (\tfrac{1}{3} \beta^2 + \tfrac{3}{5} \beta \sigma^2
        + \tfrac{5}{21} \sigma^4)
        + \delta \sigma u_t^2,
\\
\label{Ttn1rank9}
T^t &=& -\delta \sigma^2 (\beta^2 + 3 \beta \sigma^2
        + \tfrac{5}{3} \sigma^4) u_t
        - \tfrac{1}{3} \delta^3 u_t^3
\nonumber \\
    &=& \tfrac{1}{3} \delta \left(
        (\beta - 5 \sigma^2) (\beta + \sigma^2)^2 - \beta^3 \right) u_t
        - \tfrac{1}{3} \delta^3 u_t^3,
\end{eqnarray}
for $n=2$ with a density of rank $7$; and
\begin{eqnarray}
\label{Tyn1rank9-newlaw}
T^y &=& \sigma^3 (\tfrac{1}{3} \beta^3 + \tfrac{9}{10} \beta^2 \sigma^2
        + \tfrac{5}{7} \beta \sigma^4 + \tfrac{7}{36} \sigma^6)
        + \delta \sigma u_t^2,
\\
\label{Ttn1rank12}
T^t &=& -\delta \sigma^2 (\beta^3 + \tfrac{9}{10} \beta^2 \sigma^2
        + 5 \beta \sigma^4 + \tfrac{7}{4} \sigma^6) u_t
        - \tfrac{1}{3} \delta^3 u_t^3
\nonumber \\
    &=& \tfrac{1}{4} \delta \left(
        (\beta - 7 \sigma^2) (\beta + \sigma^2)^3 - \beta^4 \right) u_t
        - \tfrac{1}{3} \delta^3 u_t^3,
\end{eqnarray}
for $n=3$ and a density of rank $9$.

Inspection of (\ref{Ttn1rank6}), (\ref{Ttn1rank9}), and
(\ref{Ttn1rank12}) reveals the form of the flux
\begin{equation}
\label{Ttn-rank3nplus3}
T^t = \tfrac{\delta}{n+1} \left(
      (\beta-(2n+1) \sigma^2) (\beta + \sigma^2)^n -\beta^{n+1} \right) u_t
      - \tfrac{1}{3} \delta^3 u_t^3
\end{equation}
for arbitrary $n$.
Based on (\ref{Tyn1rank5}), (\ref{Tyn1rank7}), and (\ref{Tyn1rank9-newlaw}),
one can assume that for arbitrary $n$ the density will take the form
\begin{equation}
\label{formTyn-rank2nplus3}
T^y = \sigma^3 F(\sigma) + \delta \sigma u_t^2,
\end{equation}
where the unknown function $F(\sigma)$ is determined as follows:
 Substitute (\ref{formTyn-rank2nplus3}) and
(\ref{Ttn-rank3nplus3}) into (\ref{conslaw}) to get the ODE
\begin{equation}
\label{ODEFsigma}
\sigma^3 F^{\prime} + 3 \sigma^2 F
 = \tfrac{1}{n+1}
 \left(
  \beta^{n+1}
  - \left( \beta - (2n+1) \sigma^2 \right)(\beta + \sigma^2)^n
 \right).
\end{equation}
Use, e.g., {\em Mathematica}, to compute the general solution
of (\ref{ODEFsigma}),
\begin{equation}
\label{solFsigma}
F(\sigma) \!=\!
 \tfrac{\beta^{n+1}}{(n+1) \sigma^2} \!+\! \tfrac{c}{\sigma^3}
 \!-\! \tfrac{\beta^{n+1}}{(n+1) \sigma^2} \,
 {}_2F_1(\tfrac{1}{2}, -n; \tfrac{3}{2}; -\tfrac{\sigma^2}{\beta})
\!+\! \tfrac{2n+1}{3(n+1)} \beta^n \,
 {}_2F_1(\tfrac{3}{2}, -n; \tfrac{5}{2}; -\tfrac{\sigma^2}{\beta}),
\end{equation}
 where the integration constant $c$ can be set to zero to
avoid a constant term in $T^y$ in (\ref{formTyn-rank2nplus3}).
Substitute (\ref{solFsigma}) into (\ref{formTyn-rank2nplus3}) to get

\begin{equation}
\label{Tyn-rank2nplus3-final} \!\! T^y \!=\!
\tfrac{\beta^{n+1}}{n+1} \sigma
    \!-\! \tfrac{\beta^{n+1}}{3(n+1)} \sigma \,
    {}_2F_1(\tfrac{1}{2}, -n; \tfrac{3}{2}; -\tfrac{\sigma^2}{\beta})
\!+\! \tfrac{2n+1}{3 (n+1)} \beta^n \sigma^3 \,
 {}_2F_1(\tfrac{3}{2}, -n; \tfrac{5}{2}; -\tfrac{\sigma^2}{\beta}).
\end{equation}
Evaluation of (\ref{Tyn-rank2nplus3-final}) for $n=1, 2$, and $3$
yields (\ref{Tyn1rank5}), (\ref{Tyn1rank7}), and
(\ref{Tyn1rank9-newlaw}), respectively.

The strategies described in
Sections~\ref{compute-conslaw4} and~\ref{compute-conslaw5}, as well
as the method used in the example above can be applied to compute
conservation laws of increasingly higher ranks. In particular, a
family of densities of rank $3n+4$ (with matching fluxes) can be
obtained.  With \verb|ConservationLawsMD.m| we computed the
following density-flux pairs:
\begin{eqnarray}
\label{Tyn1r7-newest}
T^y &=& \sigma^2 u
   \left( 3 \beta + \tfrac{9}{2} \sigma^2 \right) \sigma_t
  - \delta^2 \sigma u_t^3,
\\
\label{Ttn1r8-newest}
T^t &=& -\tfrac{1}{\delta} \sigma^4
       \left( \tfrac{3}{4} \beta^2
       + \tfrac{5}{4} \beta \sigma^2
       + \tfrac{9}{16} \sigma^4 \right)
 \nonumber \\
 && + \, \delta \sigma^2
   \left( \tfrac{3}{2} \beta + \tfrac{9}{4} \sigma^2 \right)
   ( u_t^2 - 2 u u_{tt} )
 + \tfrac{1}{4} \delta^3 u_x^4,
\end{eqnarray}
for $n=1$ when requesting a density of rank 7;
\begin{eqnarray}
\label{Tyn2r10-newest}
T^y &=& \sigma^2 u
  ( 3 \beta^2 + 9 \beta \sigma^2 + 5 \sigma^4 ) \sigma_t
  - \delta^2 \sigma u_t^3,
\\
\label{Ttn2r12-newest}
T^t &=& -\tfrac{1}{\delta} \sigma^4
  \left( \tfrac{3}{4} \beta^4
  + \tfrac{5}{2} \beta^3 \sigma^2
  + \tfrac{13}{4} \beta^2 \sigma^4
  + \tfrac{19}{10} \beta \sigma^6
  + \tfrac{5}{12} \sigma^8 \right)
\nonumber \\
&& \, + \delta \sigma^2
   \left( \tfrac{3}{2} \beta^2 + \tfrac{9}{2} \beta \sigma^2
   + \tfrac{5}{2} \sigma^4 \right)
   (u_t^2 - 2 u u_{tt})
 + \tfrac{1}{4} \delta^3 u_x^4,
\end{eqnarray}
for $n=2$ and searching for a density of rank 10; and
\begin{eqnarray}
\label{Tyn3r13-newest}
\!\!\!\!\!\!\!\!\!\!\!\!
T^y &=& \sigma^2 u
 \left( 3 \beta^3 + \tfrac{27}{2} \beta^2 \sigma^2
 + 15 \beta \sigma^4 + \tfrac{21}{4} \sigma^6 \right) \sigma_t
 - \delta^2 \sigma u_t^3,
\\
\label{Ttn3r16-newest}
\!\!\!\!\!\!\!\!\!\!\!\!
T^t &\!=\!& -\tfrac{1}{\delta} \sigma^4
 \left( \tfrac{3}{4} \beta^6
  \!+\! \tfrac{15}{4} \beta^5 \sigma^2
  \!+\! \tfrac{129}{16} \beta^4 \sigma^4
  \!+\! \tfrac{75}{8} \beta^3 \sigma^6
  \!+\! \tfrac{99}{16} \beta^2 \sigma^8
  \!+\! \tfrac{123}{56} \beta \sigma^{10}
  \!+\! \tfrac{21}{64} \sigma^{12} \right)
\nonumber \\
&& \, + \delta \sigma^2
   \left( \tfrac{3}{2} \beta^3 + \tfrac{27}{4} \beta^2 \sigma^2
   + \tfrac{15}{2} \beta \sigma^4 + \tfrac{21}{8} \sigma^8 \right)
   (u_t^2 - 2 u u_{tt})
 + \tfrac{1}{4} \delta^3 u_x^4,
\end{eqnarray}
for $n=3$ with a density of rank 13.

In contrast with the previous cases, the above densities and
fluxes do not readily reveal the explicit form for either the
density or flux. However, it is obvious that the density will be of
the form
\begin{equation}
\label{formTyn-rank3nplus4}
T^y = \sigma^2 u F(\sigma) \sigma_t - \delta^2 \sigma u_t^3,
\end{equation}
where the unknown function $F(\sigma)$ which can be determined as
follows: Compute ${\mathrm D}_{y} \, T^y$ and, as usual, evaluate
the result on solutions of (\ref{magan-(3.7)-(3.8)}) to get $P$.
Require that ${\cal{E}}_{\sigma(t)} P \equiv 0$ and
${\cal{E}}_{u(t)} P \equiv 0$ which leads to ODE
\begin{equation}
\label{ODEFsigma2ndnewconslaw}
\sigma F^{\prime} + 2 F
= 6 \left( \beta + (2 n + 1) \sigma^2 \right) (\beta + \sigma^2)^{n-1}
\end{equation}
with general solution
\begin{equation}
\label{solFsigma2ndnewconslaw}
F(\sigma) =
- \tfrac{3}{(n+1) \sigma^2} \left(\beta - (2n+1)\sigma^2 \right)
(\beta + \sigma^2)^n + \tfrac{c}{\sigma^2},
\end{equation}
where $c$ is an arbitrary integration constant.
Set $c_1 = \tfrac{3\beta^{n+1}}{n+1}$ to avoid a constant term in $T^y$.
Then substitute
\begin{equation}
\label{solFsigma2ndnewconslawfinal}
F(\sigma) =
\tfrac{3}{(n+1) \sigma^2}
\Big( \beta^{n+1} - \left( \beta - (2n+1)\sigma^2 \right)
(\beta + \sigma^2)^n \Big)
\end{equation}
into (\ref{formTyn-rank3nplus4}) to get
\begin{equation}
\label{Tyn-rank3nplus4-final}
T^y = \tfrac{3 u}{(n+1)}
\Big( \beta^{n+1} - \left(\beta - (2n+1)\sigma^2 \right)
(\beta + \sigma^2)^n \Big) \sigma_t - \delta^2 \sigma u_t^3.
\end{equation}
One can readily verify that for $n=1, 2$, and $3$
density (\ref{Tyn-rank3nplus4-final}) reduces to (\ref{Tyn1r7-newest}), (\ref{Tyn2r10-newest}), and (\ref{Tyn3r13-newest}), respectively.
Finally, apply the homotopy operator to $-P$ to get the flux,

\begin{eqnarray}
\label{Ttn-rank3nplus4-final}
\!\!\!
T^t &=& - \tfrac{1}{4(n+1)^2 (2n+1)\delta}
\left( \,
3 \beta^{2(n+1)} \!+\! 6 (2n+1)\beta^{n+1} (\beta + \sigma^2)^{n+1}
\right.
\nonumber \\
\!\!\!
&& \left. - 3  \left( (4n+3) \beta - (2n+1)^2 \sigma^2 \right)
 (\beta + \sigma^2)^{2n+1}
\, \right)
\nonumber \\
\!\!\!
&& +\, \tfrac{3 \delta}{2(n+1)}
\Big(
\beta^{n+1} \!-\! \left( \beta \!-\! (2n+1) \sigma^2 \right)
  (\beta \!+\! \sigma^2)^n \Big) (u_t^2 \!-\! 2 u u_{tt})
\!+\! \tfrac{1}{4} \delta^3 u_x^4,
\end{eqnarray}
 which for $n=1, 2,$ and $3$ reduces to (\ref{Ttn1r8-newest}),
(\ref{Ttn2r12-newest}), and (\ref{Ttn3r16-newest}), respectively.
\vspace*{-2mm}
\section{Conclusions and future work}
\label{conclusions-future-work}
\vspace*{-2mm}
In this paper, Lie-point symmetries,
closed-form solutions, and conservation laws are derived for a
constitutive equation modeling stress in elastic materials, governed
by a system of nonlinear coupled PDEs (\ref{magan-(3.7)-(3.8)}). We
have determined that the Lie algebra for the model is
five-dimensional for the shearing exponent $n>0$. There are five
types of Lie symmetries: translations in time, space, and
displacement, as well as time-dependent displacement changes and a
scaling symmetry. Using the Lie symmetry method, the optimal system
of one-dimensional subalgebras is constructed.

In the second part of the paper, closed-form solutions are computed using
the optimal system of one-dimensional subalgebras.
The reductions and resulting solutions are summarized in a table.
Some of the closed-form solutions involving both variables $y$
and $t$ might help better understand the physics of the model.
In future research, one could consider appropriate initial and boundary conditions to further explore the properties of these models in the context of power-law fluids.

In this paper we also have reported seven polynomial conservation
laws for system (\ref{magan-(3.7)-(3.8)}) with arbitrary $n$. There
are likely infinitely many conservation laws.  From past
experiences, we know that PDEs with conserved densities of
increasing higher ranks usually have other interesting
``integrability" properties (see, e.g.,
\cite{hereman-goktas-mca-2024}). This prompted us to run the
Painlev\'{e} test to verify if (\ref{magan-(3.7)-(3.8)}) has the
Painlev\'e property meaning that its solutions have no worse
singularities than movable poles. System (\ref{magan-(3.7)-(3.8)})
fails the Painlev\'{e} test. We also searched for higher-order
(generalized) symmetries and a recursion operator that would connect
them. We found some polynomial generalized symmetries but no
recursion operator. These results are premature and require
additional research. In addition, we searched for special solutions
in terms of the hyperbolic functions ${\mathrm{tanh}}$ and
${\mathrm{sech}}$, as well as the Jacobi elliptic functions
${\mathrm{cn}}$ and ${\mathrm{sn}}$, without any success.
Nevertheless, we believe that (\ref{magan-(3.7)-(3.8)}) has more
structure which will be investigated in the future.
\vspace*{-2mm}
\section{Dedication}
\label{dedication}
\vspace*{-2mm}
This paper is dedicated to Prof.\ David Mason at the occasion of his
$80^{\mathrm th}$ birthday and in honor of his long career in mathematics.
WH is grateful to Prof.\ Mason for his significant contributions
to the applied sciences, in particular, in the areas of symmetry analysis,
theoretical mechanics, and fluid dynamics. Like diamonds from South
Africa, his mathematical papers are gems with excellent
(mathematical) weight, multiple facets, and great clarity. They are
cut and polished by a craftsman with skill and care. Equally
important, through his dedication, kindness, and humanity, Prof.
Mason had set an example for his students, collaborators, and for
all of us, teachers and researchers alike.

RN would like to express her deepest gratitude  to her PhD
advisor, Prof.\ Mason, for his invaluable support and guidance
throughout her academic journey. From assisting with presentation
slides to reviewing her research papers, Prof.\ Mason was always
there, generously offering his time and expertise. His commitment to
providing timely feedback, engaging in thorough discussions, and
giving personal attention had a profound impact on her development.

Through his mentorship, Prof.\ Mason played a pivotal role
in shaping RN into the academic she is today, refining her abilities
as a teacher, researcher, and mentor. With his guidance, RN secured
research funding from the National Research Foundation (NRF) for her
PhD studies and a short-term postdoctoral fellowship. His support
also helped her earn the Best Tutor Award during her time as a tutor
for his honors students. Under his mentorship, RN not only honed her
teaching skills but also grew as a mentor to her own students. Prof.
Mason's approach to student interactions shaped her understanding of
how to inspire and nurture young minds, fostering a supportive and
engaging learning environment.

Even after completing her PhD, Prof.\ Mason has remained a guiding
force, always available to offer advice on teaching, research, and
academic endeavors. His warmth and generosity were further
exemplified when he personally welcomed RN during her visit to South
Africa in 2014, where she attended a conference celebrating his 70th
birthday.
\vspace*{-2mm}
\section*{Acknowledgment}
\label{acknowledgment}
\vspace*{-2mm}
WH would like to thank Prof.\ Stephen Anco for his hospitality and
insightful discussions during a visit at Brock University. We are
grateful to the anonymous referees for their valuable comments and
suggestions.
\vspace*{-2mm}
\section*{Declaration of competing interest}
\vspace*{-2mm}
The authors declare that they have no known competing financial
interests or personal relationships that could have appeared to
influence the work reported in this paper.
\vspace*{-2mm}
\section*{Data Availability}
\vspace*{-2mm}
Data sharing is not applicable to this article as no data sets were
generated or analysed during the current study.

\end{document}